\documentclass[journal]{IEEEtran}

\newcommand{\comm}[1]{{{#1}}}
\newcommand{\revise}[1]{{{#1}}}
\newcommand{\rev}[1]{{{#1}}}

\usepackage{color, soul}

\newcounter{MYtempeqncnt}
\usepackage{subcaption}

\usepackage{amsfonts}
\usepackage{amsmath}
\usepackage{amssymb}

\usepackage{amsthm}
\newtheorem{definition}{Definition}
\newtheorem{proposition}{Proposition}

\usepackage{algorithm}
\usepackage{algpseudocode}

\DeclareMathOperator{\Unif}{Unif}

\usepackage{graphicx}
\usepackage{hyperref}

\begin{document}

\title{Two-dimensional multi-target detection: an autocorrelation analysis approach}

\author{Shay~Kreymer
        and~Tamir~Bendory%
\thanks{S. Kreymer and T. Bendory are with the School of Electrical Engineering of Tel Aviv University, Tel Aviv, Israel, e-mail: \href{mailto:shaykreymer@mail.tau.ac.il}{shaykreymer@mail.tau.ac.il}, \href{mailto:bendory@tauex.tau.ac.il}{bendory@tauex.tau.ac.il}. \comm{S.K. is supported by the Yitzhak and Chaya Weinstein Research Institute for Signal Processing.} T.B. is supported in part by NSF-BSF grant no. 2019752, the Zimin Institute for Engineering Solutions Advancing Better Lives, \rev{BSF grant no. 2020159, and ISF grant no. 1924/21}.}%
}

\markboth{IEEE Transactions on Signal Processing}%
{Kreymer \MakeLowercase{\textit{et al.}}: Two-dimensional Multi-target Detection}

\IEEEpubid{0000--0000/00\$00.00~\copyright~2022 IEEE}

\maketitle

\begin{abstract}
We consider the two-dimensional multi-target detection problem of recovering a target image from a noisy measurement that contains multiple copies of the image, each randomly rotated and translated. Motivated by the structure reconstruction problem in single-particle cryo-electron microscopy, we focus on the high noise regime, where the noise hampers accurate detection of the image occurrences. We develop an autocorrelation analysis framework to estimate the image directly from a measurement with an arbitrary spacing distribution of image occurrences, bypassing the estimation of individual locations and rotations. We conduct extensive numerical experiments, and demonstrate image recovery in highly noisy environments. The code to reproduce all numerical experiments is publicly available at~\url{https://github.com/krshay/MTD-2D}.
\end{abstract}


\section{Introduction}
\label{sec:introduction}
\IEEEPARstart{W}{e} study the multi-target detection (MTD) problem of estimating a target image~\mbox{$f:\mathbb{R}^2 \rightarrow \mathbb{R}$} from a noisy measurement that contains multiple copies of the image, each randomly rotated and translated~\cite{bendory2019multi}, \cite{lan2020multi}, \cite{marshall2020image}, \cite{bendory2021multi}, \cite{bendory2018toward}. Specifically, let~\mbox{$M: \{0, \ldots, N-1\}^2 \rightarrow \mathbb{R}$} be a measurement of the form
\begin{equation}
\label{eq:model}
M[\vec{\ell}] = \sum_{i=1}^{p} F_{\phi_i}[\vec{\ell} - \vec{\ell}_i] + \varepsilon[\vec{\ell}],
\end{equation}
where
\begin{itemize}
{\item \mbox{$\{\phi_i\}_{i=1}^{p} \sim \Unif[0, 2\pi)$} are uniformly distributed rotations;
\item \mbox{$F_{\phi_i} [\vec{\ell}] := f_{\phi_i} (\vec{\ell} / n)$} is a discrete copy of~$f$, rotated by angle~$\phi_i$ about the origin;~$n$ is a fixed integer;}
\item \mbox{$\{\vec{\ell}_i\}_{i=1}^{p} \in \{n + 1, \ldots, N-n\}^2$} are arbitrary translations;
\item $\varepsilon[\vec{\ell}]$ is i.i.d. Gaussian noise with zero mean and \mbox{variance~$\sigma^2$}.
\end{itemize}

\begin{figure}[!tb]
	\begin{subfigure}[ht]{0.30\columnwidth}
		\centering
		\includegraphics[width=\columnwidth]{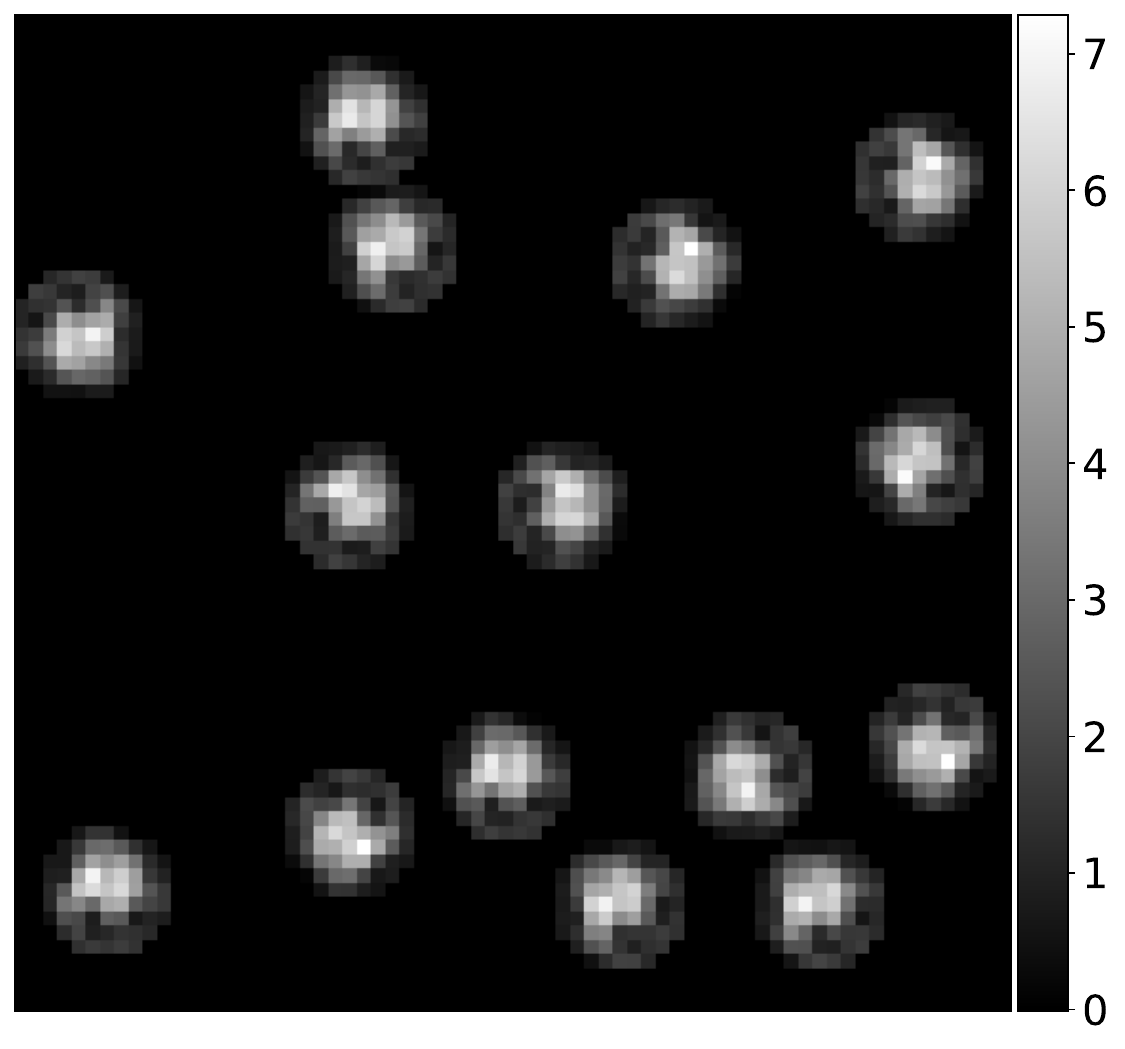}
		\caption{No noise.}
	\end{subfigure}
	\hfill
	\begin{subfigure}[ht]{0.30\columnwidth}
		\centering
		\includegraphics[width=\columnwidth]{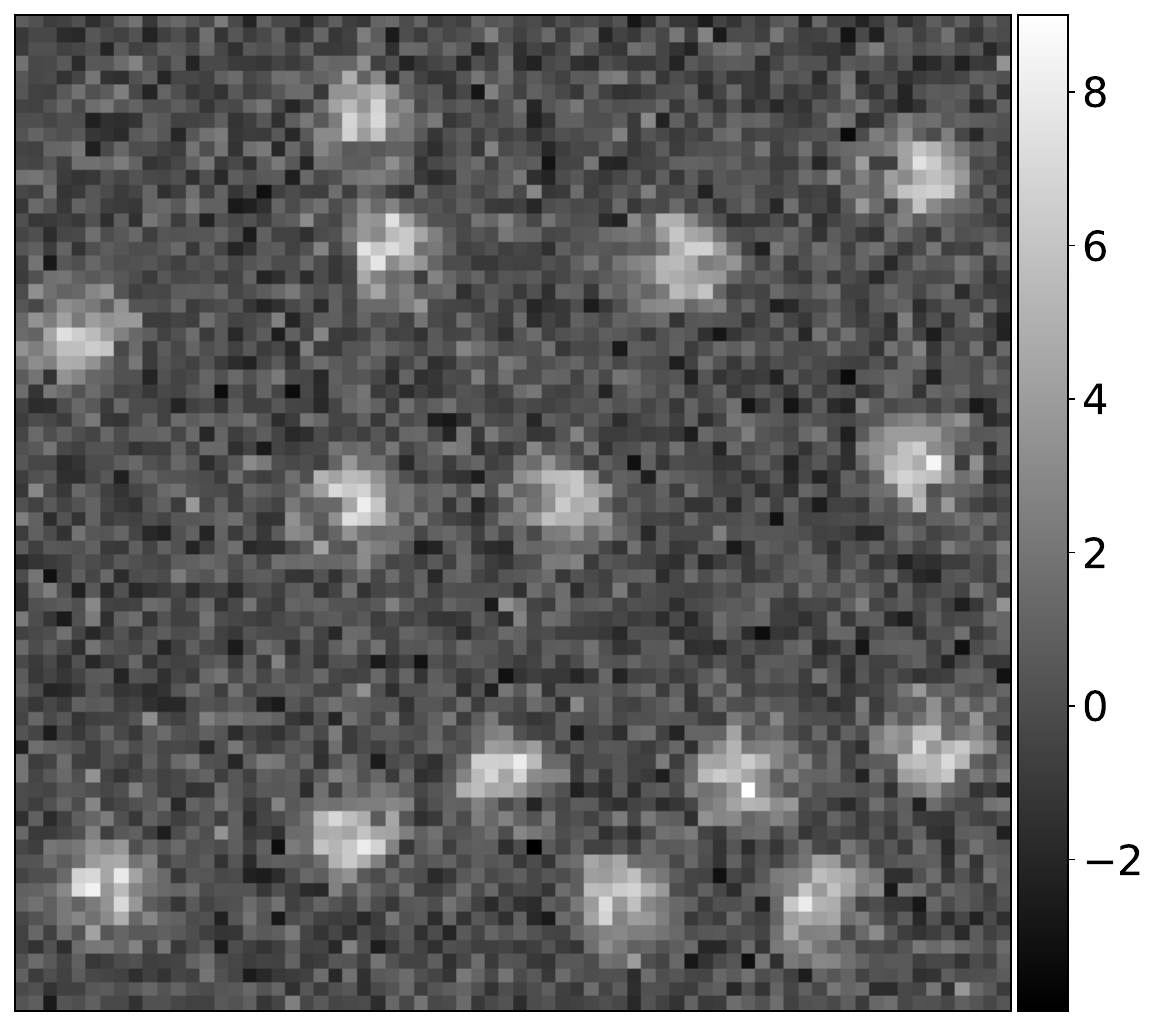}
		\caption{$\text{SNR} = 10$.}
	\end{subfigure}
	\hfill
	\begin{subfigure}[ht]{0.30\columnwidth}
		\centering
		\includegraphics[width=\columnwidth]{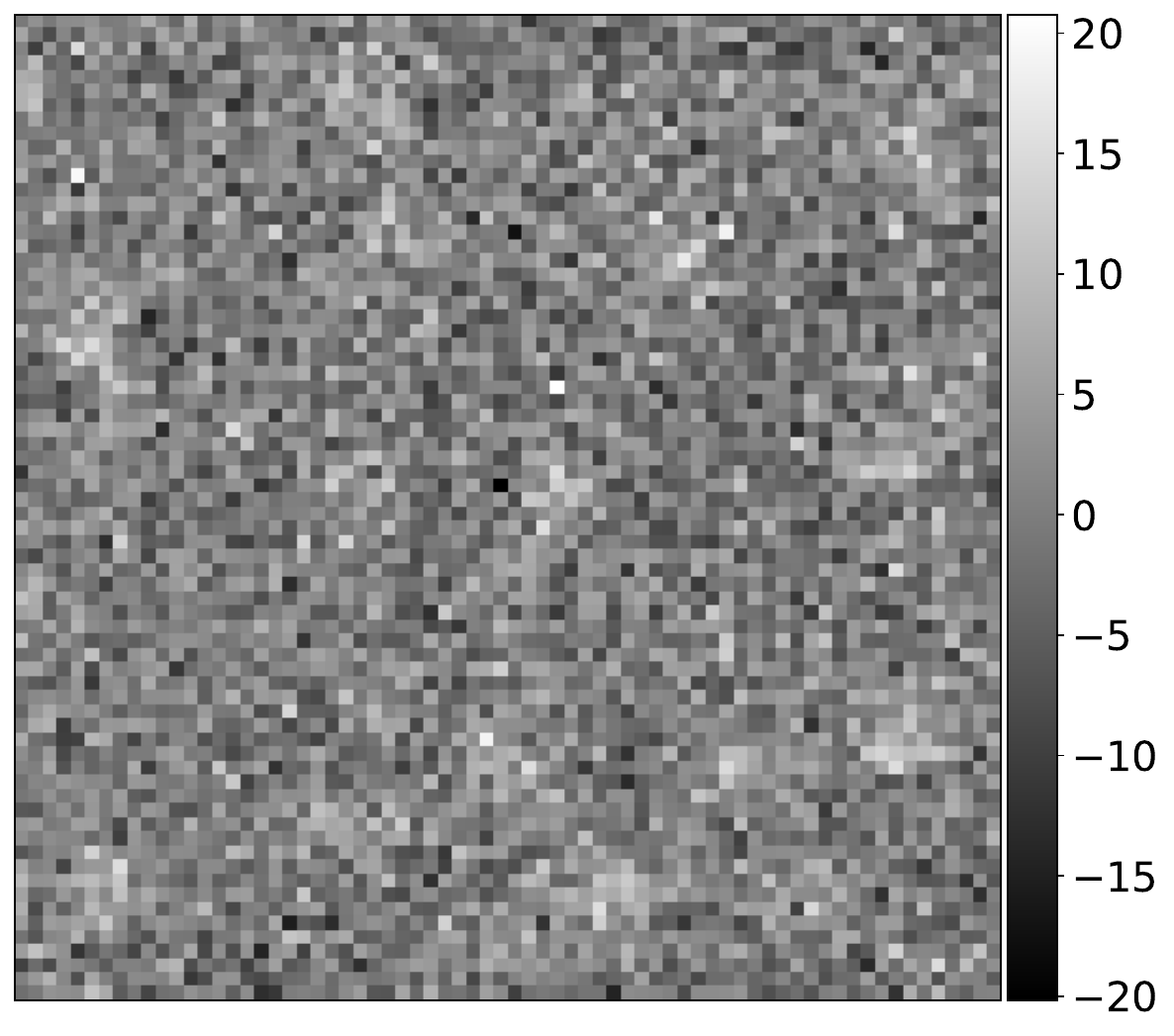}
		\caption{$\text{SNR} = 0.5$.}
	\end{subfigure}
	\caption{Three measurements at different noise levels: no noise~(left);~\mbox{$\text{SNR} = 10$} (middle);~\mbox{$\text{SNR} = 0.5$} (right). Each measurement contains multiple rotated versions of the target image in arbitrary locations. In this work, our goal is to estimate the target image directly from~\comm{the measurement}. We focus on the low SNR regime~{(e.g., panel~(c))} in which the image occurrences are swamped by the noise, and the locations and rotations of the image occurrences cannot be detected reliably.}
\label{fig:Micrographs_noise}
\end{figure}

The rotations, translations and the number of occurrences of~$f$ in~$M$ are unknown. {Importantly, since the rotations are {unknown}, it is possible to reconstruct the target image only up to a rotation.} Section~\ref{subsec:image_model} introduces the {image} model {of~$f$} in detail. Figure~\ref{fig:Micrographs_noise} presents an example of a measurement~$M$ at different signal-to-noise ratios (SNRs). We define~\mbox{$\text{SNR} := \frac{\|F\comm{_0}\|_\text{F}^2}{A \sigma^2}$}, where~$A$ is the area \comm{in pixels of~$F\comm{_0}$ (the unrotated image)}, and~$\sigma^2$ is the noise variance.

The MTD model arises in several scientific applications, such as passive radar~\cite{gogineni2017passive}, astronomy~\cite{schulz1993multiframe}, motion deblurring~\cite{levin2006blind}, and system identification~\cite{abed1997blind}. In particular, it serves as \comm{a} mathematical abstraction of the cryo-electron microscopy~(\mbox{cryo-EM}) technology for macromolecular structure determination~\cite{henderson1995potential},~\cite{nogales2016development},~\cite{bai2015cryo}. In a \mbox{cryo-EM} experiment \cite{frank2006three}, biological macromolecules suspended in a liquid solution are rapidly frozen into a thin ice layer. An electron beam then passes through the sample, and a two-dimensional tomographic projection is recorded. Importantly, the \mbox{2-D} location and \mbox{3-D} orientation of particles within the ice are random and unknown. This measurement, called \textit{micrograph}, is affected by high noise levels and the optical configuration of the microscope. This transformation is typically modeled as a convolution of the model~(\ref{eq:model}) with a point spread function, whose Fourier transform is called contrast transfer function~(CTF)~\cite{heimowitz2020reducing}, \cite{erickson1971measurement}.
\IEEEpubidadjcol

In the current analysis workflow of \mbox{cryo-EM} data \cite{bendory2020single}, \cite{scheres2012relion}, \cite{punjani2017cryosparc}, the~{\mbox{2-D}} projections are first detected and extracted from the micrograph, and later rotationally and translationally aligned to reconstruct the~{\mbox{3-D}} molecular structure. This approach fails for small molecules, which induce low contrast, and thus low SNR. This makes them difficult to detect and align~\cite{bendory2018toward}, \cite{henderson1995potential}, \cite{bendory2020single}, \cite{aguerrebere2016fundamental}, rendering current \mbox{cryo-EM} algorithmic pipeline ineffective. For example, in the limit~\mbox{$\text{SNR} \rightarrow 0$}, reliable detection of signals' locations within the measurement is impossible~\cite[Proposition~3.1]{bendory2018toward}.

The MTD model was devised in \cite{bendory2018toward} {in order to study the recovery of small molecules directly from the micrograph}, below the current detection limit of \mbox{cryo-EM}~\cite{henderson1995potential},~\cite{d2021current}. {An autocorrelation analysis technique (see Section~\ref{subsec:ac}) was implemented} to recover \mbox{low-resolution}~\mbox{3-D} structures from noiseless simulated data under a simplified model. Autocorrelation analysis consists of finding an image that best explains the empirical autocorrelations of the measurement. For any noise level, those autocorrelations can be estimated to any desired accuracy for sufficiently large~$N$. Computing the autocorrelations is straightforward and requires only one pass over the data, which is advantageous for massively large datasets, such as \mbox{cryo-EM} datasets~\cite{bendory2020single}. As such, autocorrelation analysis provides an attractive alternative to other computational methods, such as maximum likelihood estimation, which is intractable for the MTD problem~\cite{lan2020multi}. In addition, autocorrelation analysis allows to account for the CTF\comm{, under mild conditions. These conditions are summarized in Proposition~\ref{prop:ctf} in Appendix~\ref{app:ctf}.}

\begin{figure}[!tb]
	\begin{subfigure}[t]{0.45\columnwidth}
		\centering
		\includegraphics[width=\columnwidth]{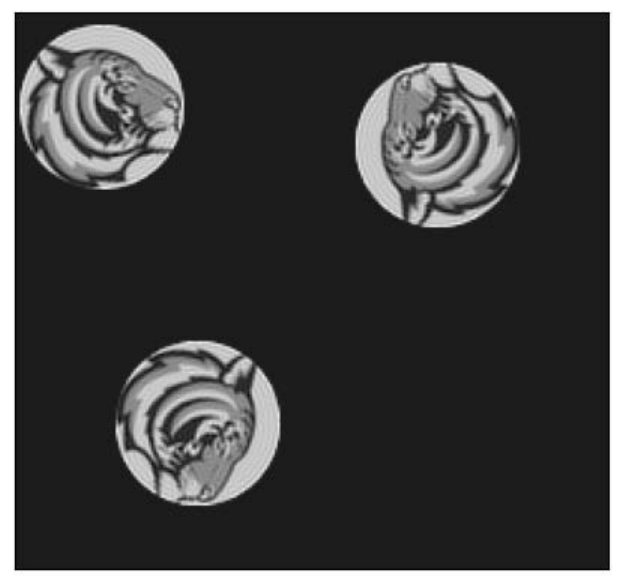}
		\caption{An example of a noiseless \mbox{well-separated} measurement with three arbitrarily rotated and translated copies of a target image.}
	\label{fig:Micrograph_shifts_a}
	\end{subfigure}
	\hfill
	\begin{subfigure}[t]{0.45\columnwidth}
	\centering
	\includegraphics[width=\columnwidth]{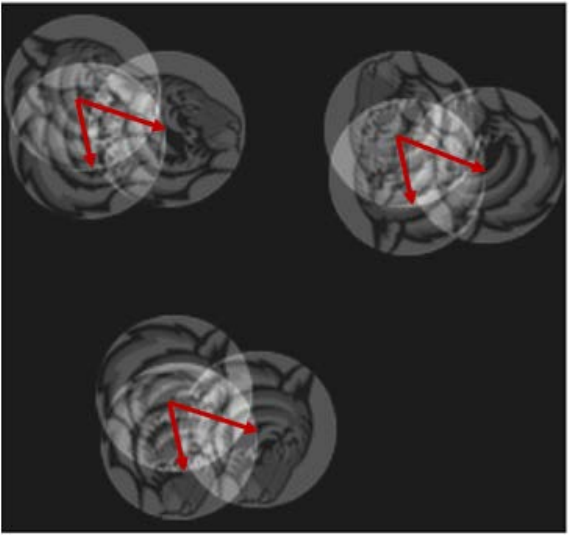}
	\caption{The measurement from (a) {and} two of its {shifted versions}.}
	\label{fig:Micrograph_shifts_b}
	\end{subfigure}

	\bigskip
	\begin{subfigure}[t]{0.45\columnwidth}
		\centering
		\includegraphics[width=\columnwidth]{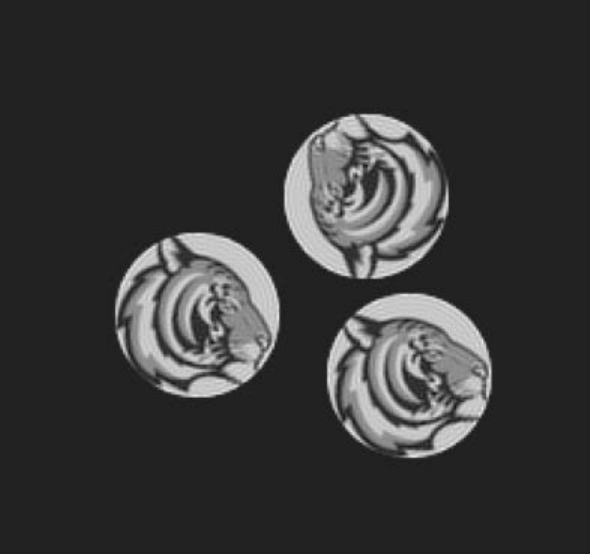}
		\caption{An example of a noiseless measurement with three arbitrarily rotated and translated copies of a target image.}
	\label{fig:Micrograph_shifts_c}
	\end{subfigure}
	\hfill
	\begin{subfigure}[t]{0.45\columnwidth}
	\centering
	\includegraphics[width=\columnwidth]{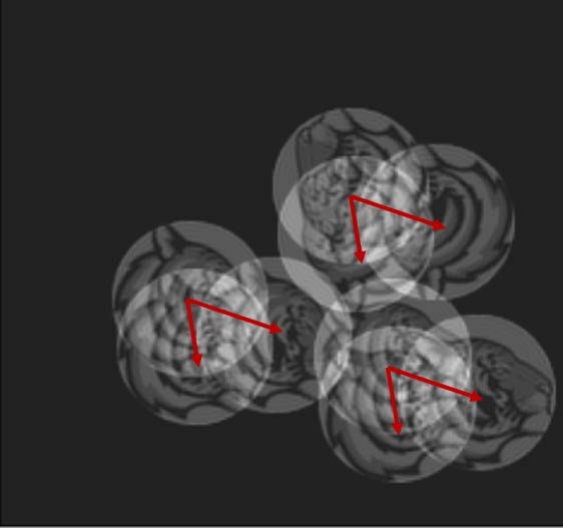}
	\caption{The measurement from (c) {and} two of its {shifted versions}.}
	\label{fig:Micrograph_shifts_d}
	\end{subfigure}

	\caption{A comparison between a \mbox{well-separated} measurement~(a) and a measurement with an arbitrary spacing distribution~(c). The third-order autocorrelation of {a} measurement~$M$ {of the form~(\ref{eq:model})} is the product of~$M$ with two shifted copies of itself (the shifts are marked by the red arrows). For \mbox{well-separated} measurements, as presented in panel~(b), any given {image occurrence} in~$M$ is only ever correlated with itself, and never with another {image occurrence}. In contrast, for the arbitrary spacing distribution case {studied in this paper} {and} illustrated in panel~(d), the third-order autocorrelation involves complicated interactions between neighboring image occurrences, and consequently between the autocorrelations of the measurement and the target image.}
	\label{fig:Micrograph_shifts}
\end{figure}

In order to further investigate the method proposed in~\cite{bendory2018toward} from analytical and computational perspectives, the MTD model was studied in~\cite{bendory2019multi} for one-dimensional signals and under the assumption that the signal occurrences are either \mbox{well-separated} or follow a Poisson distribution. In~\cite{lan2020multi}, the mathematical framework was extended to account for arbitrary spacing of one-dimensional signal occurrences. In~\cite{marshall2020image},~\cite{bendory2021multi}, the authors laid the foundations for analyzing the MTD problem in two dimensions, where the sought images are arbitrarily rotated, but still \mbox{well-separated}. This case, where each image in the measurement is separated \comm{by at least a full image diameter} from its neighbors, is presented in~Figure~\ref{fig:Micrograph_shifts_a}. Notably, when computing the entries of the measurement's autocorrelations which are smaller than the image's diameter, each image occurrence interacts only with itself (as demonstrated in~Figure~\ref{fig:Micrograph_shifts_b}), significantly simplifying the relations between the autocorrelations of the measurement and the target image.

This work extends the analysis of the MTD model and generalizes {previous works}~\cite{bendory2019multi}, \cite{lan2020multi}, \cite{marshall2020image}, \cite{bendory2021multi} by considering two-dimensional images that are both arbitrarily rotated, and arbitrarily spaced; the image occurrences are only required to not overlap. When dealing with an \emph{arbitrary spacing distribution} of the images in the measurement (see Figure~\ref{fig:Micrograph_shifts_c}), the autocorrelations of the measurement involve intricate interactions between adjacent image occurrences (see for instance~Figure~\ref{fig:Micrograph_shifts_d}), leading to complicated relations between the autocorrelations of the measurement and the \comm{sought} image.

The main contribution of this paper is in {developing an autocorrelation analysis framework for the} two-dimensional MTD problem {with} an arbitrary spacing distribution\comm{. We} define the separation functions between a pair and a triplet of adjacent image occurrences in the measurement, \comm{allowing us to} relate the autocorrelations of the measurement with those of the target image. Then, we devise an algorithm for the recovery of the target image from {a measurement}, and demonstrate a successful reconstruction in {noisy regimes}~(see Section~\ref{sec:numeric}). It is \comm{thus} a significant step towards efficiently estimating a molecular structure directly from a noisy \mbox{cryo-EM} micrograph~\cite{bendory2018toward}.

\section{Mathematical framework}
\label{sec:math}
\subsection{Image model}
\label{subsec:image_model}
We consider an image \mbox{$f: \mathbb{R}^2 \rightarrow \mathbb{R}$}, which is supported on the unit disk \mbox{$D = \{\vec{x} \in \mathbb{R}^2: |\vec{x}| \le 1\}$}. We assume that~$f$ has a finite expansion in the basis of Dirichlet Laplacian eigenfunctions; this is a standard assumption in the literature~\cite{bendory2021multi},~\cite{zhao2013fourier},~\cite{zhao2016fast}, which is akin to assuming that the image is bandlimited. This implies that we can expand the image (see Appendix~\ref{app:bandlimited}) as
\begin{equation}
\label{eq:expansion}
f(r, \theta) = \sum_{(\nu, q): \lambda_{\nu, q} \le \lambda} \alpha_{\nu, q} \psi_{\nu, q} (r, \theta), \quad \text{for } r \le 1,
\end{equation}
\comm{in polar coordinates~$(r, \theta)$}, where~$\lambda$ is called the bandlimit frequency. \comm{The expansion coefficients are denoted by~$\alpha_{\nu, q}$}, and
\begin{equation}
\label{eq:eigenfunctions}
\psi_{\nu, q}(r,\theta) = J_\nu\left( \lambda_{\nu, q} r \right) e^{i \nu \theta}, \quad \text{for } r \le 1,
\end{equation}
where~$\nu \in \mathbb{Z}_{\ge 0}$,~$J_\nu$ is the~\mbox{$\nu$-th} order Bessel function of the first kind, and~$\lambda_{\nu, q} > 0$ is the~\mbox{$q$-th} positive root of~$J_\nu$\comm{, \mbox{where~$\lambda_{\nu, q} = \lambda_{-\nu, q}$}}. This expansion is known as \comm{the} \mbox{Fourier-Bessel} expansion. \comm{The number of required coefficients is given by the sampling criterion provided in~\cite{zhao2016fast}, \cite{klug1972three}}\comm{, which is the analog of the classical Nyquist sampling rate}. For each fixed~$\nu$, we define
\begin{equation}
g_\nu (r, \theta) = \sum_{q: \lambda_{\nu, q} \le \lambda} \alpha_{\nu, q} \psi_{\nu, q} (r, \theta),
\end{equation}
so that
\begin{equation}
\label{eq:f_steerable}
f(r, \theta) = \sum_{\nu = -\nu_{\text{max}}}^{\nu_{\text{max}}} g_\nu (r, \theta),
\end{equation}
where~$\nu_{\text{max}} := \text{max}\{\nu: \lambda_{\nu, 1} \le \lambda\}$.

The basis of Dirichlet Laplacian eigenfunctions is steerable: rotating~$f$ is equivalent to modulating the expansion coefficients~$\alpha_{\nu, q}$. Specifically,~$f_\phi (r, \theta) := f(r, \theta + \phi)$ can be computed by multiplying each term in~(\ref{eq:f_steerable}) by~$e^{i \nu \phi}$:
\begin{equation}
\label{eq:steering}
f_\phi (r, \theta) = \sum_{\nu = -\nu_{\text{max}}}^{\nu_{\text{max}}} e^{i \nu \phi} g_\nu (r, \theta).
\end{equation}

As actual \mbox{cryo-EM} measurements are discretized {on} a Cartesian grid, we will focus on the analysis of a discrete version of~$f$\comm{, under the assumption of \comm{point-wise} sampling}. The discrete image~$F_\phi: \mathbb{Z}^2 \rightarrow \mathbb{R}$ is \comm{thus} defined by
\begin{equation}
F_\phi [\vec{\ell}] = f_\phi (\vec{\ell} / n), \quad \text{for } \vec{\ell} \in \mathbb{Z}^2,
\end{equation}
where~$n$ is a fixed integer that determines the sampling resolution. In our case,~$n$ is the radius of~$f$, in pixels. Since~$f_\phi$ is supported on the unit disk~$D$, it follows that~$F_\phi$~is supported~on~$\{\vec{\ell} \in \mathbb{Z}^2: |\vec{\ell}| \le n\}$.

Let~$\Psi_{\nu, q}: \mathbb{Z}^2 \rightarrow \mathbb{C}$ be the discretization of the Dirichlet Laplacian eigenfunctions~(\ref{eq:eigenfunctions})
\begin{equation}
\Psi_{\nu, q} [\vec{\ell}] = \psi_{\nu, q} (\vec{\ell} / n),
\end{equation}
and let~$\hat{\Psi}_{\nu, q}: \mathbb{Z}^2 \rightarrow \mathbb{C}$ be the discrete Fourier transform (DFT) of~$\Psi_{\nu, q}$
\begin{equation}
\label{eq:Psi_k}
\hat{\Psi}_{\nu, q} [\vec{k}] = \sum_{\vec{\ell} \in \mathbb{Z}^2} \Psi_{\nu, q} [\vec{\ell}] e^{-2 \pi i \vec{\ell} \cdot \vec{k} / (4n)}.
\end{equation}
With this notation,
\begin{align}
\label{eq:rot_F}
F_\phi [\vec{\ell}] &= \sum_{(\nu, q): \lambda_{\nu, q} \le \lambda} \alpha_{\nu, q} \Psi_{\nu, q} [\vec{\ell}] e^{i \nu \phi} \nonumber\\&= \sum_{\nu=-\nu_{\text{max}}}^{\nu_{\text{max}}} \Bigg(\sum_{q:\lambda_{\nu, q} \le \lambda} \alpha_{\nu, q} \Psi_{\nu, q}[\vec{\ell}]\Bigg) e^{i\nu\phi},
\end{align}
and the DFT of~$F_\phi$,~$\hat{F}_\phi: \mathbb{Z}^2 \rightarrow \mathbb{C}$, is given by
\begin{equation}
\hat{F}_{\phi} [\vec{k}] = \sum_{\vec{\ell} \in \mathbb{Z}^2} F_\phi [\vec{\ell}] e^{-2 \pi i \vec{\ell} \cdot \vec{k} / (4n)}.
\end{equation}
Using~(\ref{eq:Psi_k}) and the linearity of the DFT, we conclude
\begin{equation}
\hat{F}_{\phi} [\vec{k}] = \sum_{(\nu, q): \lambda_{\nu, q} \le \lambda} \alpha_{\nu, q} \hat{\Psi}_{\nu, q} [\vec{k}] e^{i \nu \phi}.
\end{equation}

\subsection{Autocorrelation analysis}
\label{subsec:ac}
The autocorrelation of order~$q$ of a random signal~\mbox{$z \in \mathbb{R}^{N \times N}$} is defined as
\begin{equation}
A_z^q[\vec{\ell}_1, \ldots, \vec{\ell}_{q-1}] := \mathbb{E}_z\Big[\frac{1}{N^2} \sum_{\vec{i} \in \mathbb{Z}^2} z[\vec{i}] z[\vec{i} + \vec{\ell}_1] \cdots z[\vec{i} + \vec{\ell}_{q-1}]\Big],
\end{equation}
where~$\vec{\ell}_1, \ldots, \vec{\ell}_{q-1}$ are integer shifts. {Indexing} out of bounds is zero-padded, that is,~\mbox{$z[\vec{i}] = 0$} out of the range~\mbox{$\{0, \ldots, {N-1}\} \times \{0, \ldots, {N-1}\}$}. {In this work}, we use the first three autocorrelations which are explicitly given by
\begin{align}
\label{eq:ac1}
A_z^1 &= \mathbb{E}_z \left[\frac{1}{N^2} \sum_{\vec{i} \in \mathbb{Z}^2} z\left[\vec{i}\right] \right], \\
\label{eq:ac2}
A_z^2\left[\vec{\ell}\right] &= \mathbb{E}_z \left[\frac{1}{N^2} \sum_{\vec{i} \in \mathbb{Z}^2} z\left[\vec{i}\right] z\left[\vec{i} + \vec{\ell}\right] \right], \\
\label{eq:ac3}
A_z^3\left[\vec{\ell}_1, \vec{\ell}_2\right] &= \mathbb{E}_z \left[\frac{1}{N^2} \sum_{\vec{i} \in \mathbb{Z}^2} z\left[\vec{i}\right] z\left[\vec{i} + \vec{\ell}_1\right] z\left[\vec{i} + \vec{\ell}_2\right] \right].
\end{align}
As~$N^2$ grows indefinitely, the empirical autocorrelations of~$z$ almost surely (a.s.) converge to the population autocorrelations of~$z$
\begin{equation}
\lim_{N \rightarrow \infty} \frac{1}{N^2} \sum_{\vec{i} \in \mathbb{Z}^2} z[\vec{i}] z[\vec{i} + \vec{\ell}_1] \cdots z[\vec{i} + \vec{\ell}_{q-1}] \stackrel{\text{a.s.}}{=}A_z^q[\vec{\ell}_1, \ldots, \vec{\ell}_{q-1}],
\end{equation}
\comm{by the law of large numbers}.

Our goal is to relate the autocorrelations of the measurement with the discretized target image~$F$. {In particular}, the first-order autocorrelation is defined as
\begin{equation}
\label{eq:AM1}
A_{M}^1 := \frac{1}{N^2} \sum_{\vec{i} \in \mathbb{Z}^2} M[\vec{i}].
\end{equation}
This is the mean of the measurement. The second-order autocorrelation of~$M$, \mbox{$A_{M}^2: \mathbb{Z}^2 \rightarrow \mathbb{R}$}, is defined by
\begin{equation}
\label{eq:AM2}
A_{M}^2 [\vec{\ell}_1] := \frac{1}{N^2} \sum_{\vec{i} \in \mathbb{Z}^2} M[\vec{i}] M[\vec{i} + \vec{\ell}_1],
\end{equation}
and the third-order autocorrelation~\mbox{$A_{M}^3: \mathbb{Z}^2 \times \mathbb{Z}^2 \rightarrow \mathbb{R}$} by
\begin{equation}
\label{eq:AM3}
A_{M}^3 [\vec{\ell}_1, \vec{\ell}_2] := \frac{1}{N^2} \sum_{\vec{i} \in \mathbb{Z}^2} M[\vec{i}] M[\vec{i} + \vec{\ell}_1] M[\vec{i} + \vec{\ell}_2].
\end{equation}

We now introduce the \mbox{well-separated} model of the MTD problem. Then, we address the arbitrary spacing distribution case of the MTD problem, which is the main contribution of this work.

\subsection{MTD with \mbox{well-separated} images}
\label{subsec:well_separated}
We first discuss the \mbox{well-separated} case of the 2-D MTD problem, which was studied in~\cite{marshall2020image} and~\cite{bendory2021multi}. In this case, we assume that each image in the measurement~$M$ is separated by at least a full image diameter from its neighbors. Specifically, we assume that
\begin{equation}
\label{eq:sep}
|\vec{\ell}_{i_1} - \vec{\ell}_{i_2}| > 4n, \quad \text{ for all } i_1 \ne i_2.
\end{equation}

Figure~\ref{fig:Micrograph_shifts_a} presents an example of a measurement obeying the separation condition~(\ref{eq:sep}). To compute the third-order autocorrelation~(\ref{eq:AM3}), we compute the product of~$M$ with its two shifts. As demonstrated in Figure~\ref{fig:Micrograph_shifts_b}, for~\mbox{$\vec{\ell}$-s} in the range
\begin{equation}
\label{eq:set_L}
{\mathcal{L} = \{0, \ldots, {2n}\} ^ 2},
\end{equation}
any given occurrence of~$F$ in~$M$ is only ever correlated with itself, and never with another occurrence.

To understand the relation between the autocorrelations of the measurement and the target image~$F$, we first need to define the target image's autocorrelations. Specifically,~$S_1^{\alpha}$, the mean of the image, is given by
\begin{equation}
\label{eq:S1}
S_1^{\alpha} := \frac{1}{(2n + 1)^2} \sum_{\vec{\ell} \in \mathbb{Z}^2} F[\vec{\ell}],
\end{equation}
\comm{where \comm{the superscript~$\alpha$ emphasizes the dependence on the Fourier-Bessel expansion coefficients of~$F$}.}
The \emph{rotationally-averaged} second-order autocorrelation of~$F$, $S_2^{\alpha} : \mathbb{Z}^2 \rightarrow \mathbb{R}$, is given by
\begin{equation}
\label{eq:S2}
S_2^{\alpha}[\vec{\ell}_1] := \frac{1}{(2n + 1)^2} \frac{1}{2 \pi} \int_0^{2 \pi} \sum_{\vec{\ell} \in \mathbb{Z}^2} F_\phi [\vec{\ell}] F_\phi [\vec{\ell} + \vec{\ell}_1] d\phi,
\end{equation}
and~$S_3^{\alpha}: \mathbb{Z}^2 \times \mathbb{Z}^2 \rightarrow \mathbb{R}$, the rotationally-averaged third-order autocorrelation of~$F$, is given by
\begin{multline}
\label{eq:S3}
S_3^{\alpha} [\vec{\ell}_1, \vec{\ell}_2] := \frac{1}{(2n+1)^2} \frac{1}{2 \pi} \int_0^{2 \pi} \sum_{\vec{\ell} \in \mathbb{Z}^2} F_\phi [\vec{\ell}] \times \\ \times F_\phi [\vec{\ell} + \vec{\ell}_1] F_\phi [\vec{\ell} + \vec{\ell}_2] d\phi.
\end{multline}
The autocorrelations of the target image,~$S_1^{\alpha}$,~$S_2^{\alpha} $ and~$S_3^{\alpha}$ are supported on~\mbox{$\mathcal{J} :=\{-2n, \ldots, 2n\}^2 \subset \mathbb{Z}^2$}, and are invariant to rotations.

In~\cite{marshall2020image}, it was shown that under the separation condition~(\ref{eq:sep}), for any fixed level of noise~$\sigma^2$, density~$\gamma$ and image radius~$n$, in the limit~\mbox{$N \rightarrow \infty$} we have that
\begin{align}
\label{eq:well_separated_1st}
A_{M}^1 &\stackrel{\text{a.s.}}{=} \gamma S_1^{\alpha}, \\
\label{eq:well_separated_2nd}
A_{M}^2 [\vec{\ell}_1] &\stackrel{\text{a.s.}}{=} \gamma S_2^{\alpha} [\vec{\ell}_1] + \sigma^2\delta[\vec{\ell}_1], \\
\label{eq:well_separated_3rd}
A_{M}^3 [\vec{\ell}_1, \vec{\ell}_2] &\stackrel{\text{a.s.}}{=} \gamma S_3^{\alpha} [\vec{\ell}_1, \vec{\ell}_2] \nonumber \\&+ \gamma S_1^{\alpha} \sigma^2 (\delta[\vec{\ell}_1] + \delta[\vec{\ell}_2] + \delta[\vec{\ell}_1 - \vec{\ell}_2]),
\end{align}
for~$\vec{\ell}_1, \vec{\ell}_2 \in \mathcal{L}$ {{(defined in~(\ref{eq:set_L}))}}, where
\begin{equation}
\label{eq:delta}
\delta[\vec{\ell}] = \begin{cases} 1 \text{ if } \vec{\ell} = \vec{0}, \\ 0 \text{ otherwise}, \end{cases}
\end{equation}
is the Kronecker delta function. Here,~$\gamma$ is the density of the target images in the measurement and is {defined} by
\begin{equation}
\label{eq:gamma}
\gamma = p \frac{(2n + 1)^2}{N^2}.
\end{equation}

As such,~\mbox{(\ref{eq:well_separated_1st}) -~(\ref{eq:well_separated_3rd})} relate the autocorrelations of the measurement with the target image~$F$. {Previous works~\cite{bendory2019multi}, \cite{lan2020multi}, \cite{marshall2020image}, \cite{bendory2021multi} demonstrated successful signal and image estimations \comm{from the autocorrelations of the measurement}.} Importantly, the aforementioned relations between the autocorrelations of~$M$ and~$F$ do not directly depend on the {location of individual image occurrences} in the measurement, but only through the density parameter~$\gamma$. Therefore, detecting the image occurrences is not a prerequisite for image recovery, and thus image recovery is possible even in very low SNR regimes.

\subsection{MTD with an arbitrary spacing distribution}
\label{subsec:arbitrary_spacing_distribution}
We now discuss the case of arbitrary spacing distribution of image occurrences in~$M$. In contrast to the \mbox{well-separated} model, the autocorrelations of~$M$ may now involve correlating distinct occurrences of~$F$, in various relative positions and rotations. This is illustrated in Figures~\ref{fig:Micrograph_shifts_c} and~\ref{fig:Micrograph_shifts_d}. There, a measurement with an arbitrary spacing distribution and its two shifted versions are presented to illustrate the computation of the third-order autocorrelation~(\ref{eq:AM3}). In this case, the autocorrelation of the measurement is a sum of the autocorrelation of the image with itself ({as in the \mbox{well-separated} case~(\ref{eq:well_separated_1st}) -~(\ref{eq:well_separated_3rd})}), with additional \mbox{cross-terms} contributed by its close neighbors.

{Specifically}, the autocorrelations of the measurement are related to the target image not only through its autocorrelations, but also through the distribution of its occurrences in the measurement. To characterize this distribution, we define the pair separation and triplet separation functions.

\begin{definition}[Pair separation function]
The pair separation function (PSF)~$\xi$ is defined as
\begin{equation}
\label{eq:PSF}
\xi[\vec{\ell}_1] := \frac{1}{p} \sum_{i=1}^p \sum_{\substack{j=1 \\ j \ne i}}^p \delta[\vec{\ell}_i - \vec{\ell}_j - \vec{\ell}_1],
\end{equation}
where~$p$ is the number of copies of the target image in the measurement, and~$\{\vec{\ell}_i\}_{i=1}^p$ are the arbitrary translations, as defined in~(\ref{eq:model}).
\end{definition}
The PSF~$\xi\comm{[\vec{\ell}_1]}$ is the probability that an occurrence of~$F$ in~$M$ has a neighboring {image} \comm{positioned at~$\vec{\ell}_1$ with respect to its center}. Similarly to the~PSF, we can define the average interaction of a triplet of image occurrences as follows:
\begin{definition}[Triplet separation function]
The triplet separation function (TSF)~$\zeta$ is defined as
\begin{equation}
\label{eq:TSF}
\zeta[\vec{\ell}_1, \vec{\ell}_2] := \frac{1}{p^2} \sum_{i=1}^p \sum_{\substack{j_1=1 \\ j_1 \ne i}}^p \sum_{\substack{j_2=1 \\ j_2 \ne i}}^p \delta[\vec{\ell}_i - \vec{\ell}_{j_1} - \vec{\ell}_1] \cdot \delta[\vec{\ell}_i - \vec{\ell}_{j_2} - \vec{\ell}_2],
\end{equation}
where~$p$ and~$\{\vec{\ell}_i\}_{i=1}^p$ are as defined in~(\ref{eq:PSF}).
\end{definition}
The TSF~$\zeta\comm{[\vec{\ell}_1, \vec{\ell}_2]}$ is the probability that an occurrence of~$F$ in~$M$ has neighboring {images} that \comm{are positioned at both~$\vec{\ell}_1$ and~$\vec{\ell}_2$ with respect to its center}.

\revise{
Using brute force algorithms, \comm{the complexity of computing the PSF and the TSF} is, respectively,~$\mathcal{O}(p^2)$ and~$\mathcal{O}(p^3)$. To accelerate computations, a fast nearest neighbor search method~\cite{maneewongvatana2002analysis} is utilized to reconstruct the set of image locations in the measurement,~$\{\vec{\ell}_i\}_{i = 1}^p$, as a binary trie. Then, for each occurrence, only a constant number of nearest neighbors are scanned and counted. In practice, this algorithm reduces computation time by a few orders of magnitude.
}

Recall that in the arbitrary spacing distribution case (the main focus of this work), autocorrelations of the target image with its close neighbors also contribute to the autocorrelations of the measurement. To this end, we need to define the average interaction of an image with its rotated copies. Let {us define the rotationally-averaged second-order autocorrelation~\mbox{$S_{2, \text{pair}}^{\alpha}: \mathbb{Z}^2 \rightarrow \mathbb{R}$} as follows}
\begin{multline}
\label{eq:S2_neigh}
S_{2, \text{pair}}^{\alpha}[\vec{\ell}_1] := \frac{1}{(2n + 1)^2} \Big(\frac{1}{2 \pi}\Big)^2 \int_0^{2 \pi} \int_0^{2 \pi} \sum_{\vec{\ell} \in \mathbb{Z}^2} F_\phi [\vec{\ell}] \times \\ \times F_\eta [\vec{\ell} + \vec{\ell}_1] d\phi d\eta,
\end{multline}
where the two terms are rotated independently about the origin. \comm{We further define the third-order autocorrelation of a pair of target images,~\mbox{$S_{3, \text{pair}}^{\alpha}: \mathbb{Z}^2 \times \mathbb{Z}^2 \rightarrow \mathbb{R}$},
\begin{multline}
\label{eq:S3_neigh}
S_{3, \text{pair}}^{\alpha} [\vec{\ell}_1, \vec{\ell}_2] := \frac{1}{(2n + 1)^2} \Big(\frac{1}{2 \pi}\Big)^2 \int_0^{2 \pi} \int_0^{2 \pi} \sum_{\vec{\ell} \in \mathbb{Z}^2} F_\phi [\vec{\ell}] \times \\ \times F_\phi [\vec{\ell} + \vec{\ell}_1] F_\eta [\vec{\ell} + \vec{\ell}_2] d\phi d\eta,
\end{multline}
where one of the instances is rotated independently relative to the two others, and of a triplet of target images,~\mbox{$S_{3, \text{trip}}^{\alpha}: \mathbb{Z}^2 \times \mathbb{Z}^2 \rightarrow \mathbb{R}$},
\begin{multline}
\label{eq:S3_trip}
S_{3, \text{trip}}^{\alpha} [\vec{\ell}_1, \vec{\ell}_2] := \frac{1}{(2n + 1)^2} \Big(\frac{1}{2 \pi}\Big)^3 \int_0^{2 \pi} \int_0^{2 \pi} \int_0^{2 \pi} \sum_{\vec{\ell} \in \mathbb{Z}^2} F_\phi [\vec{\ell}] \times \\ \times F_\eta [\vec{\ell} + \vec{\ell}_1] F_\iota [\vec{\ell} + \vec{\ell}_2] d\phi d\eta d\iota,
\end{multline}
where all three instances are rotated independently.} As before, the autocorrelations of~$F$ are supported on~\mbox{$\mathcal{J} = \{-2n, \ldots, 2n\}^2$}.

Based on these definitions, the autocorrelations for the arbitrary spacing distribution case are summarized in~\mbox{Proposition~\ref{prop:asd}}. The proof is given in Appendix~\ref{app:arbitrary_spacing_distribution_model}.
\begin{proposition}
\label{prop:asd}
For any fixed level of noise~$\sigma^2$, density~$\gamma$, image radius~$n$, in the limit~\mbox{$N \rightarrow \infty$}, the autocorrelations for the two dimensional arbitrary spacing distribution case are given by
\begin{align}
\label{eq:1st-order-asd}
A_{M}^1 &\stackrel{\text{a.s.}}{=} \gamma S_1^{\alpha},\\
\label{eq:2nd-order-asd}
A_{M}^2[\vec{\ell}_1] {}&\stackrel{\text{a.s.}}{=} \gamma S_2^{\alpha}[\vec{\ell}_1] \nonumber \\& {+}\:\gamma \comm{A^{\alpha}(\xi) [\vec{\ell}_1]} \nonumber \\& {+}\:\sigma^2\delta[\vec{\ell}_1],\\
\label{eq:3rd-order-asd}
A_{M}^3[\vec{\ell}_1, \vec{\ell}_2] {}&\stackrel{\text{a.s.}}{=} \gamma  S_3^{\alpha}[\vec{\ell}_1, \vec{\ell}_2] \nonumber \\ & {+}\: \comm{\gamma B^{\alpha}(\xi) [\vec{\ell}_1, \vec{\ell}_2] + \gamma C^{\alpha}(\zeta)[\vec{\ell}_1, \vec{\ell}_2]} \nonumber \\ & {+}\: \gamma S_1^{\alpha} \sigma^2 (\delta[\vec{\ell}_1] + \delta[\vec{\ell}_2]+\delta[\vec{\ell}_1 - \vec{\ell}_2]),
\end{align}
where~\mbox{$\mathcal{L}_{\text{neighbors}} := \{{-4n+2}, \ldots, {4n-1}\}^2$}, and~\mbox{$\vec{\ell}_1, \vec{\ell}_2 \in \mathcal{L}$} {{(defined in~(\ref{eq:set_L}))}}. \comm{The terms~$A^{\alpha}(\xi)$, $B^{\alpha}(\xi)$, and~$C^{\alpha}(\zeta)$ are linear functions of the PSF and {the} TSF, and their explicit expressions are\revise{, respectively, {provided} in}~(\ref{eq:A}), (\ref{eq:B}) and~(\ref{eq:C}), in Appendix~\ref{app:arbitrary_spacing_distribution_model}.}
\end{proposition}

Compared to the \mbox{well-separated} case, the contribution of the target images that violate the separation condition is summarized in \comm{three terms,~$A^{\alpha}(\xi)$, $B^{\alpha}(\xi)$, and~$C^{\alpha}(\zeta)$}. They involve the second- and third-order autocorrelations of the image~$F$ with its neighbors,~$S_{2, \text{pair}}^{\alpha}$, $S_{3, \text{pair}}^{\alpha}$ and~$S_{3, \text{trip}}^{\alpha}$, defined above. For the \mbox{well-separated} case, based on~(\ref{eq:sep}), those terms are equal to zero, {and} the autocorrelations reduce to~\mbox{(\ref{eq:well_separated_1st}) -~(\ref{eq:well_separated_3rd})}.

Similarly to the \mbox{well-separated} case,~(\ref{eq:1st-order-asd}) -~(\ref{eq:3rd-order-asd}) relate the autocorrelations of the measurement with those of the target image. {Here, the relations also depend on the unknown PSF and TSF. Crucially, accurate estimation of these functions is not essential}: as the individual locations and rotations associated with individual image occurrences, these are also nuisance variables.

\subsection{Computing the autocorrelations and gradients}
\label{subsec:calc_acs}
To compute the autocorrelations presented in Sections~\ref{subsec:well_separated} and~\ref{subsec:arbitrary_spacing_distribution}, we use their Fourier representation. We leave the technical details of this section to Appendix~\ref{app:deriv}, and state here the concluding expressions.

Let~$\mathcal{V}$ denote the set of all the pairs~$(\nu, q)$ in the expansion~(\ref{eq:expansion}). We define the column vector~$\omega_{\vec{k}, \phi} \in \mathbb{C}^\mathcal{V}$ by~$(\omega_{\vec{k}, \phi})_{\nu, q} = \hat{\Psi}_{\nu, q} [\vec{k}] e^{i\nu \phi}$, and the column vector~$\alpha \in \mathbb{C}^\mathcal{V}$ by~$(\alpha)_{\nu, q} = \alpha_{\nu, q}$; the latter encodes the parameters {describing} the target image. With this notation,~$\hat{F}_\phi [\vec{k}]$ can be expressed compactly as~$\hat{F}_\phi [\vec{k}] = \alpha^\top \omega_{\vec{k}, \phi}$. In addition, let~$\mathcal{V}_0$ denote the set of all the pairs~$(0, q)$. We define the column vector~ \mbox{${\omega_0}_{\vec{k}} \in \mathbb{C}^{\mathcal{V}_0}$} by~$({\omega_0}_{\vec{k}})_{q} = \hat{\Psi}_{0, q} [\vec{k}]$, and the column vector~$\alpha_0 \in \mathbb{C}^{\mathcal{V}_0}$ by~$(\alpha_0)_{q} = \alpha_{0, q}$. It follows that
\begin{equation}
\hat{S}_2^{\alpha} [\vec{k}_1] = \frac{1}{4\nu_{\max} (2n + 1)^2} \sum_{\nu = 0}^{4\nu_{\max}} |{\alpha}^\top \omega_{\vec{k}_1, \phi_{\nu_2}}|^2,
\end{equation}
\begin{equation}
\hat{S}_{2, \text{pair}}^{\alpha} [\vec{k}_1] = \frac{1}{(2n + 1)^2} \Big({\alpha_0}^\top {\omega_0}_{-\vec{k}_1}\Big) \Big({\alpha_0}^\top {\omega_0}_{\vec{k}_1}\Big),
\end{equation}
\begin{multline}
\hat{S}_3^{\alpha} [\vec{k}_1, \vec{k}_2] = \frac{1}{6\nu_{\max} (2n + 1)^2} \sum_{\nu=0}^{6\nu_{\max}-1} \Big(\alpha^\top \omega_{\vec{k}_1, \phi_{\nu_3}}\Big) \times \\ \times\Big(\alpha^\top \omega_{\vec{k}_2, \phi_{\nu_3}}\Big)  \Big(\alpha^\top \omega_{-\vec{k}_1 - \vec{k}_2, \phi_{\nu_3}}\Big),
\end{multline}\comm{
\begin{multline}
\hat{S}_{3,\text{pair}}^{\alpha}[\vec{k}_1, \vec{k}_2] = \frac{1}{4\nu_{\max} (2n+1)^2} {\alpha_0}^\top {\omega_0}_{\vec{k}_2} \times \\ \times \sum_{\nu=0}^{4\nu_{\max}-1} \Big(\alpha^\top \omega_{\vec{k}_1, \phi_{\nu_2}}\Big) \Big(\alpha^\top \omega_{-\vec{k}_1 - \vec{k}_2, \phi_{\nu_2}}\Big),
\end{multline}
\begin{multline}
\hat{S}_{3, \text{trip}}^{\alpha} [\vec{k}_1, \vec{k}_2] = \frac{1}{(2n+1)^2} \Big({\alpha_0}^\top {\omega_0}_{\vec{k}_1}\Big) \Big({\alpha_0}^\top {\omega_0}_{\vec{k}_2}\Big)\times \\ \times \Big({\alpha_0}^\top {\omega_0}_{- \vec{k}_1 - \vec{k}_2}\Big),
\end{multline}}
where~$\phi_{\nu_2} := 2\pi\nu/(4\nu_{\text{max}})$,~$\phi_{\nu_3} := 2\pi\nu/(6\nu_{\text{max}})$, and~$\nu_{\text{max}}$ is defined in~(\ref{eq:f_steerable}). In this form, {computing} the gradients of the autocorrelations with respect to~$\alpha$ is straightforward. The gradients are given in~(\ref{eq:gS2}) -~(\ref{eq:gS3_trip}) in Appendix~\ref{app:deriv}.

\subsection{Image recovery {from autocorrelations}}
\label{subsec:recovery}
Recall that the second stage of the autocorrelation analysis framework entails recovering the image from the measurement's autocorrelations. Following~\cite{bendory2019multi}, \cite{lan2020multi}, \cite{marshall2020image}, \cite{bendory2021multi}, and based on~\mbox{(\ref{eq:1st-order-asd}) -~(\ref{eq:3rd-order-asd})}, we formulate a non-convex least squares problem for estimating the coefficients vector~$\alpha$ that represents the target image~\comm{$F$} from the autocorrelations of the measurement~$M$:
\begin{multline}
\label{eq:optimization}
\min_{\alpha, \gamma > 0, \xi, \zeta} w_1 \Bigl(A_{M}^1 - \gamma S_1^{\alpha}\Bigr)^2 \\+ w_2 \sum_{\vec{\ell}_1 \in \mathcal{L}}\Bigl(A_{M}^2[\vec{\ell}_1] -  \gamma (S_2^{\alpha}[\vec{\ell}_1] + \comm{A^{\alpha}(\xi)[\vec{\ell}_1]})  -\sigma^2\delta[\vec{\ell}_1]\Bigr)^2 \\+ w_3\sum_{\{\vec{\ell}_1, \vec{\ell}_2\} \in \mathcal{L} \times \mathcal{L}} \Bigl(A_{M}^3[\vec{\ell}_1, \vec{\ell}_2] - \gamma (S_3^{\alpha}[\vec{\ell}_1, \vec{\ell}_2] + \comm{B^{\alpha}(\xi) [\vec{\ell}_1, \vec{\ell}_2]} \\\comm{+ C^{\alpha}(\zeta) [\vec{\ell}_1, \vec{\ell}_2]} + S_1^{\alpha} \sigma^2 (\delta[\vec{\ell}_1] + \delta[\vec{\ell}_2]+\delta[\vec{\ell}_1 - \vec{\ell}_2]))\Bigr)^2,
\end{multline}
where the weights are set \mbox{to~$w_1 = 1/2$}, \mbox{$w_2 = 1/2n_2$}, \mbox{$w_3 = 1/2n_3$}, where~$n_2=4n^2$ is the number of elements in the set~$\mathcal{L}$, and~$n_3~=~16n^4$ is the number of elements in the \mbox{set~$\mathcal{L} \times \mathcal{L}$}\comm{, such that each \comm{term in}~\mbox{(\ref{eq:1st-order-asd}) -~(\ref{eq:3rd-order-asd})} is \comm{equally weighted}}. This is a non-convex (polynomial of degree~6) optimization problem, and thus there is no guarantee to converge to a global optimum. {Nevertheless}, similarly to previous papers on MTD, our numerical results (Section~\ref{sec:numeric}) suggest that standard gradient-based methods succeed in recovering~$\alpha$ \comm{from \comm{only} a few random initial guesses}.

Recall that~$\xi$ and~$\zeta$ are treated throughout this work as nuisance variables---we do not aim at estimating them, as long as the estimation of the target image~$F$ (or, more precisely, its expansion coefficients) succeeds. Those functions are unknown and depend on the density~$\gamma$, which is also unknown. \comm{Fortunately}, numerical experiments indicate that for an {accurate estimate of}~$\gamma$, the image can be accurately recovered with \emph{approximated}~$\xi$ and~$\zeta$.

The approximation of~$\xi$ and~$\zeta$ is done by \comm{simulating the image occurrences' locations~$\{\vec{\ell}_i'\}_{i = 1}^p$ according to~(\ref{eq:model})} for a given value of~$\gamma$. Specifically, for a given~$\gamma$, we calculate~$p$, the number of target images in the measurement, from~(\ref{eq:gamma}), and then simulate~$p$ \comm{random} locations, such that there are no overlaps. From these simulated locations, we calculate~$\xi'$ and~$\zeta'$ {numerically according} to~(\ref{eq:PSF}) and~(\ref{eq:TSF}), respectively. While these functions do not describe accurately {our specific} measurement~$M$ and its actual separation relations, we noticed that when~$\gamma$ is accurately estimated, we succeed in estimating~$\alpha$ using the approximated~$\xi'$ and~$\zeta'$.

Moreover, numerical experiments (see Section~\ref{subsec:gamma_exp}) {indicate that a successful estimate of $\gamma$ is achieved} when we \revise{initialize} the optimization procedure with a \comm{close enough} initial~$\gamma'$ and corresponding~$\xi'$ and~$\zeta'$. Thus, the proposed scheme consists of two \comm{stages}. {First, we} initialize the optimization {problem}~(\ref{eq:optimization}) with an arbitrary~$\gamma'$ and corresponding approximated~$\xi'$ and~$\zeta'$ from \comm{simulated locations} as explained above, and {minimize~(\ref{eq:optimization}) until a satisfactory estimate of~$\gamma$ is achieved}. \comm{This step \revise{might be repeated several times} until~$\gamma$ does not change significantly between iterations.} {\comm{Second}, we} initialize the optimization procedure again with the estimated~$\gamma$ and the updated estimates of \comm{the PSF and TSF, denoted by}~$\xi_{\text{final}}$ and~$\zeta_{\text{final}}$, to ultimately estimate~$\alpha$ by minimizing~(\ref{eq:optimization}) for fixed~$\xi_{\text{final}}$ and~$\zeta_{\text{final}}$. The scheme is described in Algorithm~\ref{alg:image_recovery}.

\begin{algorithm}[!tb]
\caption{Image recovery in the arbitrary spacing distribution case}
\label{alg:image_recovery}
\begin{algorithmic}[1]
\State Set initial values for~$\gamma$ and~$\alpha$:~$\gamma'$ and~$\alpha'$.
\comm{\Repeat
\State \label{stg:2} \comm{Simulate the image occurrences' locations~$\{\vec{\ell}_i'\}_{i = 1}^p$} according to (\ref{eq:model}) with~$\gamma'$, and calculate initial guesses~$\xi'$ and~$\zeta'$ according to~(\ref{eq:PSF}) and~(\ref{eq:TSF}).
\State \label{stg:3} Minimize~(\ref{eq:optimization}) until a satisfactory estimate~$\gamma'$ is achieved, while~$\xi = \xi'$ and~$\zeta = \zeta'$ remain fixed.
\Until the value of $\gamma'$ does not change significantly}
\State Simulate the image occurrences' locations~$\{\vec{\ell}_i''\}_{i = 1}^p$ with the estimated~$\gamma'$ from stage~\ref{stg:3}, and calculate~$\xi_{\text{final}}$ and~$\zeta_{\text{final}}$ according to~(\ref{eq:PSF}) and~(\ref{eq:TSF}).
\State Minimize~(\ref{eq:optimization}) until a satisfactory estimate of~$\alpha$ is achieved, while~$\xi = \xi_{\text{final}}$ and~$\zeta = \zeta_{\text{final}}$ remain fixed.
\end{algorithmic}
\end{algorithm}

{The described scheme is much more efficient than minimizing~(\ref{eq:optimization}) directly since we do not run the optimization over \comm{the variables}~$\xi$ and~$\zeta$. {Although we only approximate~$\xi$ and~$\zeta$, we observe only a slight degradation in accuracy compared} to an ideal case of known~$\xi$ and~$\zeta$~(see Section~\ref{subsec:conv_exp})}.

\comm{
\subsection{Computational complexity}
\label{subsec:computational_complexity}
The estimation of the vector of coefficients~$\alpha$ consists of three main \revise{ingredients}. First, we calculate the empirical autocorrelations of the measurement~$M$ according to~(\ref{eq:AM1}) - (\ref{eq:AM3}). The computational complexity of this stage is of~$\mathcal{O}(n^4 N^2)$, since we are interested in\comm{~$\mathcal{O}(n^4)$} shifts in the set~$\mathcal{L} \times \mathcal{L}$, where~$\mathcal{L}$ is defined in~(\ref{eq:set_L}), and computing the autocorrelations consists of~$\mathcal{O}(N^2)$ multiplications and summations. \comm{Since~$n \ll N$, we compute each autocorrelation directly, \revise{and not using FFT}, as described in Section~\ref{subsec:ac}}. Crucially, this stage is done only once. \revise{Second}, we aim to estimate the separation functions. The running time of estimating~$\xi$ and~$\zeta$ using the algorithm explained in Section~\ref{subsec:arbitrary_spacing_distribution} is negligible; this computation may be repeated a few times at the beginning of the optimization procedure. In addition, since we are not interested in the accurate separation functions, the estimation might be done using only a fraction of the data. \revise{Third}, at each iteration of the minimization of~(\ref{eq:optimization})\revise{,} we calculate the rotationally-averaged autocorrelations of the target image,~$S_1^{\alpha}$, $S_2^{\alpha}$, $S_{2, \text{pair}}^\alpha$, $S_3^{\alpha}$, $S_{3, \text{pair}}^\alpha$ and~$S_{3, \text{trip}}^\alpha$, and their gradients with respect to~$\alpha$, as detailed in Section~\ref{subsec:calc_acs}. Following~\mbox{\cite[Proposition~4.2]{bendory2021multi}}, this \comm{step} can be done in~$\mathcal{O}(n^5)$ operations, such that the total computational complexity of the minimization procedure is of~$\mathcal{O}(n^5 V)$, where~$V$ is the number of iterations. \revise{Since~\rev{\mbox{$N \gg n$}, and empirically~$N \gg V$ (in our experiments, a few hundred iterations suffice to achieve accurate recoveries)}}, overall estimating the vector of coefficients~$\alpha$ can be done in~\mbox{$\mathcal{O}(n^4 N^2 + n^5 V) = \mathcal{O}(n^4 N^2)$} operations.
}

\begin{figure*}[!tb]
	\begin{subfigure}[t]{0.32\textwidth}
		\centering
		\includegraphics[width=\columnwidth]{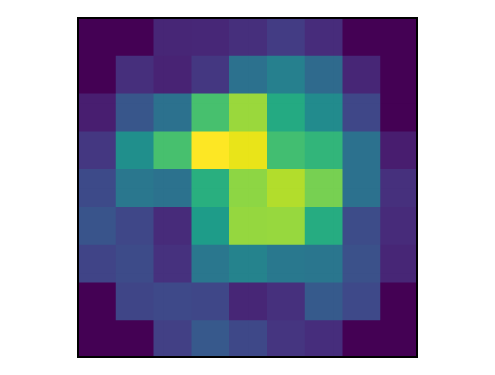}
		\caption{Target image.}
	\end{subfigure}
	\hfill
	\begin{subfigure}[t]{0.32\textwidth}
		\centering
		\includegraphics[width=\columnwidth]{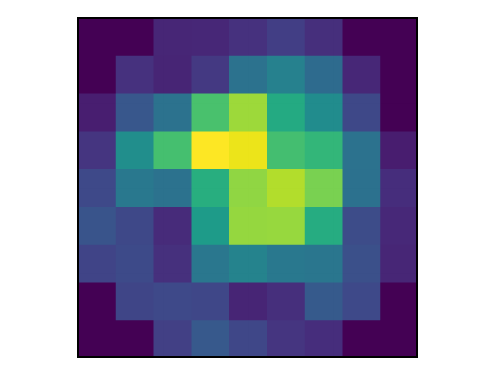}
		\caption{Estimated image with~$\text{SNR} = 10$.}
	\end{subfigure}
	\hfill
	\begin{subfigure}[t]{0.32\textwidth}
	\centering
		\includegraphics[width=\columnwidth]{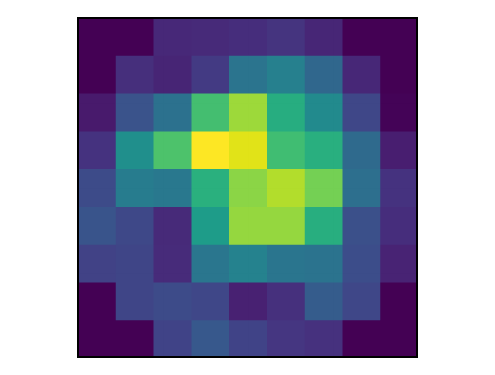}
		\caption{Estimated image with~$\text{SNR} = 0.5$.}
	\end{subfigure}

	\caption{Recovery of the target image (a) from measurements at different noise levels using Algorithm~\ref{alg:image_recovery}. {The relative error is~$0.012$ for~\mbox{$\text{SNR} = 10$}, and~$0.047$ for~\mbox{$\text{SNR} = 0.5$}}.}
	\label{fig:recoveries}
\end{figure*}

\begin{figure*}[!tb]
	\begin{subfigure}[t]{0.5\textwidth}
		\centering
		\includegraphics[width=\columnwidth]{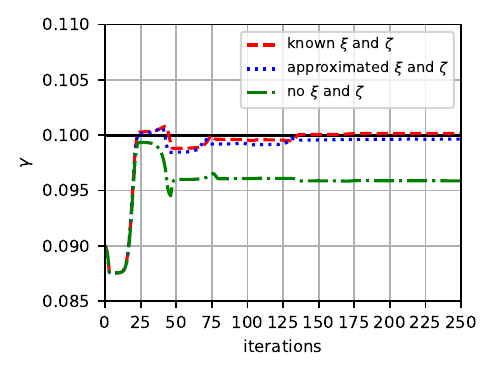}
		\caption{Initial density of~\mbox{$\gamma' = 0.09$}.}
	\label{fig:gamma_experiment_009}
	\end{subfigure}
	\hfill
	\begin{subfigure}[t]{0.5\textwidth}
		\centering
		\includegraphics[width=\columnwidth]{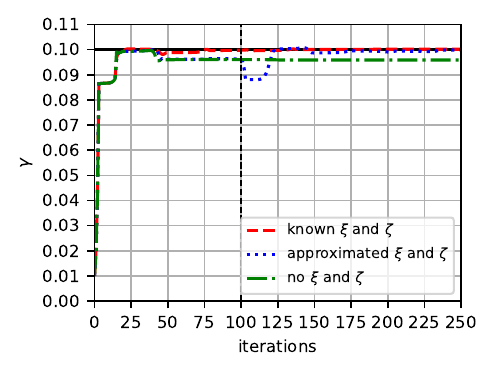}
		\caption{Initial density of~\mbox{$\gamma' = 0.01$}.}
	\label{fig:gamma_experiment_001}
	\end{subfigure}
\caption{Examples of an estimation of the density~\mbox{$\gamma^* = 0.1$} (marked by horizontal black line) from initial guesses of~\mbox{$\gamma' = 0.09$}~(left) and~\mbox{$\gamma' = 0.01$}~(right) by minimizing~(\ref{eq:optimization}) for three different cases: (1) the PSF~$\xi$ and TSF~$\zeta$ are known; (2)~approximating the PSF and TSF with an initial density~$\gamma'$; (3)~treating the MTD problem as \mbox{well-separated} and assuming no PSF or TSF. The estimate of~$\gamma$ through the iterations of the optimization procedure is presented. For~\mbox{$\gamma' = 0.01$}, the approximation of~$\gamma$ was repeated after 100 iterations (marked by the vertical dashed
black line).}
\label{fig:gamma_experiment}
\end{figure*}
\section{Numerical experiments}
\label{sec:numeric}
In this section, we present numerical results for the recovery procedure described in Section~\ref{subsec:recovery}. {The optimization problem~(\ref{eq:optimization}) was minimized using the~\mbox{Broyden-Fletcher-Goldfarb-Shanno}~(BFGS) algorithm, while ignoring the positivity constraint on~$\gamma$ (namely, treating it as an unconstrained problem)}.

{To take} the in-plane rotation symmetry into account, we {measure the estimation error by}
\begin{equation}
\label{eq:err_z}
\text{relative error}_{\alpha} := \min_{\phi \in [0, 2\pi)} \frac{\|\alpha^* - \alpha_{\phi}\|_2}{\|\alpha^*\|_2},
\end{equation}
where~$\alpha^*$ is the true vector of expansion coefficients, and~$\alpha_{\phi}$ is the vector of coefficients of the estimated image, rotated by angle~$\phi$. The relative error of {estimating} the density~$\gamma$ is defined by
\begin{equation}
\label{eq:err_gamma}
\text{relative error}_{\gamma} := \frac{|\gamma^* - \gamma|}{\gamma^*},
\end{equation}
where~$\gamma^*$ is the true density, and~$\gamma$ is the estimated density.

The code to reproduce all experiments is publicly available at~\url{https://github.com/krshay/MTD-2D}.

\subsection{Recovery from noisy measurements}
\label{subsec:recovery_exp}
In Figure~\ref{fig:recoveries} we present a successful recovery of a target image from noisy measurements using Algorithm~\ref{alg:image_recovery}. The target image is of \mbox{radius~$n = 4 \text{ pixels}$}, and it is expanded using its first~34 \mbox{Fourier-Bessel} coefficients. We consider a {dataset that} consists of~1000 {{measurements,}} each of \mbox{size~$7000 \times 7000 \text{ pixels}$}, generated according to~(\ref{eq:model}) {with density~$\gamma = 0.1$}, at different noise levels. The noise levels are visualized in Figure~\ref{fig:Micrographs_noise}. As expected, the recovery error {degrades} as the noise level {increases}. The {relative error} is~$0.012$ for~\mbox{$\text{SNR} = 10$}, and~$0.047$ for~\mbox{$\text{SNR} = 0.5$}.

\begin{figure*}[!tb]
	\begin{subfigure}[t]{0.5\textwidth}
		\centering
		\includegraphics[width=\columnwidth]{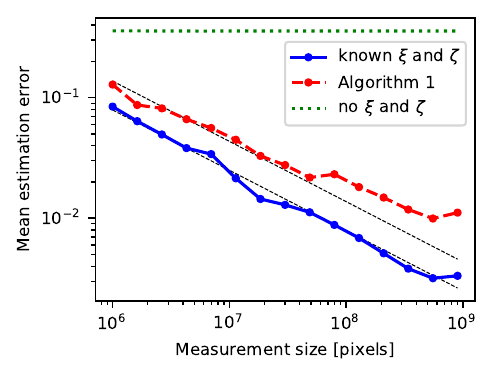}
\caption{Mean estimation error as a function of the measurement size.}
\label{fig:err_size_experiment}
	\end{subfigure}
\hfill
\begin{subfigure}[t]{0.5\textwidth}
		\centering
		\includegraphics[width=\columnwidth]{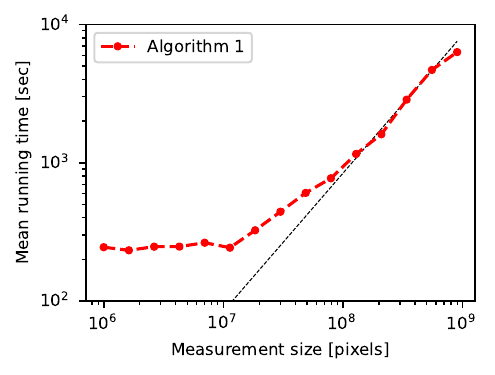}
\caption{Mean running time as a function of the measurement size.}
\label{fig:time_size}
	\end{subfigure}
	\caption{\rev{The left panel shows the mean estimation error of recovering the vector of coefficients~$\alpha$, as a function of the measurement size~$N^2$, by: (1)~Minimization of~(\ref{eq:optimization}), assuming known PSF and TSF; (2)~Algorithm~\ref{alg:image_recovery} (with approximated PSF and TSF); (3)~Minimization of~(\ref{eq:optimization}), assuming no PSF and TSF (treating the measurement as \mbox{well-separated}). The black dashed line illustrates a slope of~$-1/2$, i.e., the recovery error for case~(1) decays as~$1/\sqrt{N^2}$. The right panel shows the corresponding running time for Algorithm~\ref{alg:image_recovery}. The black dashed line illustrates a slope of~1, implying a linear increase in computation time with the measurement size~$N^2$ for large enough~$N$, agreeing with Section~\ref{subsec:computational_complexity}.}}
\end{figure*}

\subsection{Estimating the density parameter~$\gamma$}
\label{subsec:gamma_exp}
Recall that the first stages of Algorithm~\ref{alg:image_recovery} consist of successfully estimating the density~$\gamma$. In particular, we observed that a successful estimation of~$\gamma$ is possible even with an inaccurate~PSF and TSF. To demonstrate it, we present in Figure~\ref{fig:gamma_experiment_009} an example of estimation of the density~$\gamma$ from a noisy measurement with true density~\mbox{$\gamma^* = 0.1$}\comm{, and an initial guess of~$\gamma' = 0.09$}. The measurement's size is~{\mbox{$N = 25000 \text{ pixels}$}}, with~\mbox{$\text{SNR} = 0.5$}. {The target images in the measurement are of radius~\mbox{$n = 2 \text{ pixels}$}, and are expanded using their first~10~\mbox{Fourier-Bessel} coefficients}. We report the estimated~$\gamma$ through the iterations of the optimization procedure, for three different cases. In the first case, we assume that the PSF and TSF are known, and estimate~$\gamma^*$ accurately up to an error of~$0.0016$. Of course, in practice those functions are unknown. Next, we \comm{simulate the locations of the target images \comm{assuming}~$\gamma' = 0.09$,} and use the approximated PSF and TSF {in the optimization, while they} remain fixed along the iterations. Remarkably, the estimation of~$\gamma$ in this case is successful as well, with an estimation error of~$0.0036$. Finally, we try to estimate the density assuming the measurement is \mbox{well-separated}, without success; the attained recovery error is~$0.041$. {The last result} indicates that {the terms in~(\ref{eq:2nd-order-asd}), (\ref{eq:3rd-order-asd}) that encapsulate the contribution of neighboring images are crucial for the success of the recovery procedure, and cannot be ignored.}

\comm{
We repeated the same experiment with an initial guess of~\mbox{$\gamma' = 0.01$}, far from the ground truth; the results are presented in Figure~\ref{fig:gamma_experiment_001}. For this case, two repetitions of stages~\ref{stg:2} and~\ref{stg:3} of Algorithm~\ref{alg:image_recovery} were required, and we report a recovery error of~$0.0013$. As in the previous experiment, for the case of known PSF and TSF, the recovery error is~$0.0016$, and the recovery error when we assume that the measurement is well-separated is~$0.041$, which \revise{re-emphasizes} the significance of the separation functions~$\xi$ and~$\zeta$ in the arbitrary spacing distribution case. This experiment indicates that our algorithmic scheme succeeds even \revise{without a good} approximation of~$\gamma$.
}

\subsection{Recovery error as a function of the measurement size}
\label{subsec:conv_exp}
Figure~\ref{fig:err_size_experiment} presents {recovery error} as a function of the measurement size. We consider noiseless measurements {in different sizes} with an arbitrary spacing distribution. {The measurements are generated according to~(\ref{eq:model}) with density~$\gamma = 0.1$. The target images in the measurements are of radius~\mbox{$n = 2 \text{ pixels}$}, and are expanded using their first~10~\mbox{Fourier-Bessel} coefficients}. We examine three cases. In the first case, we assume the PSF and TSF are known. In the second case, {we apply Algorithm~\ref{alg:image_recovery} that uses approximated separation functions}. In the third case, we falsely assume our measurement is \mbox{well-separated}, i.e., we do not take the effect of the {PSF and TSF} into account. In all cases, we try to estimate the vector of coefficients~$\alpha$ {from} {10} random initial guesses, and calculate the estimation error {for the estimate whose final objective function is minimal}. We then calculate the mean over 50 trials.

Not surprisingly, {when the PSF and TSF are known we get a {small recovery} error}. For the first case, the error {decays} \mbox{as~$1 / \sqrt{N^2}$}, where~$N^2$ is the total number of entries in the measurement. This is the same estimation rate as if the translations and rotations were known {(that is, the estimation rate of averaging over i.i.d. Gaussian variables)}. {We get the same estimation rate for Algorithm~\ref{alg:image_recovery} {when}~$N^2$ {is} below~$10^8$, corresponding to relative error greater than approximately 1\%. {For larger~$N^2$, the error curve stagnates because of the inaccurate estimation of~$\xi$ and~$\zeta$}, thus limiting the possible estimation error of the algorithm.} Notably, we do not achieve successful recovery for the third case, regardless of the measurement size, which is {in correspondence with} the numerical results {for} the~\mbox{one-dimensional} case~\cite{lan2020multi}. This emphasizes the importance of the extension {of the MTD model} from the \mbox{well-separated} case to the arbitrary spacing distribution case.

\rev{
Figure~\ref{fig:time_size} presents the running time of Algorithm~\ref{alg:image_recovery} as a function of the measurement size. As expected from the computational complexity analysis of Section~\ref{subsec:computational_complexity}, for a large enough measurement size~$N^2$, the running time of Algorithm~\ref{alg:image_recovery} increases linearly with~$N^2$.
}

\subsection{Recovery error as a function of SNR}
\label{subsec:recovery_noise}
\begin{figure}[!tb]
	\begin{subfigure}[ht]{\columnwidth}
		\centering
		\includegraphics[width=\columnwidth]{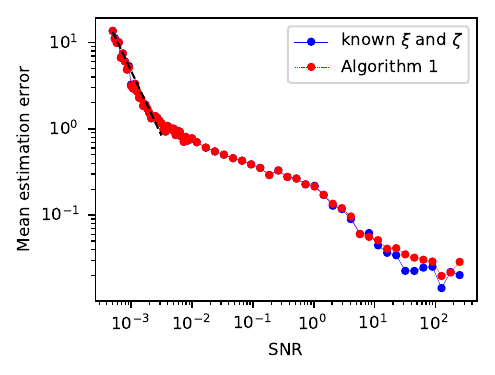}
	\end{subfigure}
	\caption{{Mean} estimation error {of recovering the} vector of coefficients~$\alpha$, {as a function of SNR}, by: (1)~Minimization of~(\ref{eq:optimization}), assuming known PSF and TSF; (2)~Algorithm~\ref{alg:image_recovery} (with approximated PSF and TSF).}
\label{fig:err_noise_experiment}
\end{figure}

Computing the $q$-th order autocorrelation consists of the product of $q$ noisy terms, and thus the variance of its estimation scales \mbox{as~$\sigma^{2q} / N$}. Under the assumption that the locations of the images are known (but the rotations are not), it was shown~\cite{abbe2018estimation}, \cite{bandeira2020optimal}, \cite{perry2019sample}, \cite{bandeira2017estimation}, \cite{romanov2021multi} that the sample complexity {of estimating the image} in the low SNR regime scales \mbox{as~$\sigma^{3}$}, or, equivalently, \mbox{as~$\text{SNR}^{{-1.5}}$}. {Figure~\ref{fig:err_noise_experiment} presents recovery errors} as a function of {the} SNR. We consider measurements with different {SNRs, and~\mbox{$N = 7000 \text{ pixels}$}}. {The target images in the measurements are of radius~\mbox{$n = 2 \text{ pixels}$}, and are expanded using their first~10~\mbox{Fourier-Bessel} coefficients. \comm{We used~{3} random initializations, and calculated the mean over~40 trials}}. {In} the low SNR regime, {the error indeed scales as $\text{SNR}^{-1.5}$, although the locations of the images are unknown}. Notably, for low SNRs {the performance gap between the methods is negligible}. In this regimes, the dominant factor is the noise, and the inaccuracy in the separation functions is inconsequential. For high SNRs, we notice a {small performance gap between the methods}, as was also noted in~{Figure~\ref{fig:err_size_experiment}}.

\subsection{Comparison with an \mbox{oracle-deconvolution} method}
\label{subsec:comparison_oracle}
\begin{figure}[!tb]
	\begin{subfigure}[ht]{\columnwidth}
		\centering
		\includegraphics[width=\columnwidth]{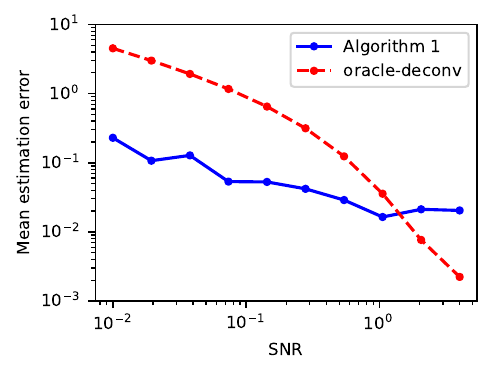}
	\end{subfigure}
	\caption{Mean estimation error of recovering \revise{a rotationally-symmetric} target image~$F$, as a function of SNR, by: (1)~Algorithm~\ref{alg:image_recovery}; (2)~the \mbox{oracle-deconvolution} method described in Section~\ref{subsec:comparison_oracle}. \comm{Evidently, the deconvolution-based method fails at low SNR.}}
\label{fig:comparison_oracle}
\end{figure}

\comm{To stress the \revise{ability} of Algorithm~\ref{alg:image_recovery} to handle high noise levels, we compare its performance against an \mbox{oracle-deconvolution} method; the results are presented in Figure~\ref{fig:comparison_oracle}. The \mbox{oracle-deconvolution} method aims to find the locations of the image occurrences in the measurement \revise{as the peaks of the convolution of the measurement with the target image}. The peaks were found using the function \mbox{\emph{photutils.detection.find\textunderscore peaks}} from the Photutils Python package~\cite{bradley2020astropy}, while forcing the peaks to be separated by at least an image diameter. Once the locations are identified (an estimate of the shifts~$\{\ell_i\}_{i = 1}^p$ in~(\ref{eq:model})), the image occurrences are extracted from the measurement and averaged. Importantly, \revise{this algorithm receives as an input} the underlying image~$F$ and the number of image occurrences~$p$, and thus the name "oracle." To avoid searching over different rotations (\revise{in order to} find the best match), we consider \revise{only in this} experiment \mbox{rotationally-symmetric} target images, so that the convolution is done only once.}

In the experiment, we consider measurements \comm{in} different SNRs \comm{with}~\mbox{$N = 20000 \text{ pixels}$}. The target images in the measurements are of radius~\mbox{$n = \rev{3} \text{ pixels}$}, and are expanded using their first~10~\mbox{Fourier-Bessel} coefficients. We used~\rev{2} random initializations, and calculated the mean over~10 trials. As expected, for high SNR regimes, the method of oracle-deconvolution deconvolution performs better than Algorithm~\ref{alg:image_recovery}. However, \comm{in} low SNR regimes, where detecting the locations of the target images becomes harder, the method fails \comm{miserably}.

\section{Conclusion}
This paper is motivated by the effort of reconstructing small~\mbox{3-D} molecular structures {using \mbox{cryo-EM}}, below the current detection limit. The main contribution of this study is expanding {the} analysis to account for an arbitrary spacing distribution of {images}, corroborated by extensive numerical experiments. \revise{In the future, we hope to extend the {framework} of this paper} to deal with the more general case of \mbox{\mbox{cryo-EM}---where} the images are different tomographic projections of a \mbox{three-dimensional} molecular structure taken from unknown viewing directions~\cite{bendory2018toward}---with an arbitrary spacing distribution between the {projection images}.

In this work, we focused on the homogeneous MTD problem, i.e., where all target images are {identical}, up to a rotation. A possible {{future study} may consider the} heterogeneous case, where the {image occurrences} are drawn either from a discrete or a continuous distribution {of possible images, and the goal is to recover the entire set of images}.

\rev{In} addition, since we acquire many copies of the target image, one can imagine that as~$N$ increases, the sampling spacing \revise{may be increased} as well. \revise{Based on \mbox{algebraic-geometry} results~\cite{bandeira2017estimation}, a similar analysis was conducted for the simpler problem of~\mbox{one-dimensional} discrete multi-reference alignment~\cite{bendory2021super}}. \revise{Extending~\cite{bendory2021super} to the MTD model might allow reducing the sampling rate of \mbox{cryo-EM} datasets, and thus alleviate the computational burden---\rev{a significant drawback of the current implementation}.} Designing other inference techniques, such as maximum likelihood estimators~\cite{lan2020multi}, \rev{\cite{kreymer2021approximate}}, for the two-dimensional MTD problem is also an interesting research thread.

In recent years, {{a various deep learning techniques} has been used to solve inverse imaging} problems, often surpassing the performance achieved {by} analytical methods~\cite{lucas2018using}, with recent endeavors in the~\mbox{cryo-EM} {literature}~\cite{gupta2020cryogan}, \cite{zhong2021cryodrgn}. Initial experiments~(not presented here) were conducted in attempt to solve the MTD problem by \emph{learning from the autocorrelations} of the measurement, {showing} promising results for {estimating} the density~$\gamma$ {and for providing a good initial estimate for the sought target image}. This indicates that learning {techniques have} the potential to robustly estimate the density, {and perhaps additional parameters}, and by that to efficiently bypass the first three stages of Algorithm~\ref{alg:image_recovery}. \rev{We expect that this strategy will improve robustness and reduce the computational complexity}.

\section*{Acknowledgment}
The authors are grateful to Ti-Yen Lan, Nicholas Marshall, and Amit Singer for insightful discussions. The authors would like to thank Nir Sharon for providing computational resources\comm{, and for the anonymous reviewers for their insightful comments.}


\appendices
\section{The effect of the contrast transfer function}
\label{app:ctf}

Let us consider a measurement~$y \in \mathbb{R}^{{N \times N}}$ given by
\begin{equation}
\label{eq:ctf_model}
y = h \ast M,
\end{equation}
where~$M \in \mathbb{R}^{N \times N}$ is as defined in~(\ref{eq:model}), $h \in \mathbb{R}^{{N \times N}}$ is a point spread function, and~$\ast$ is a circular convolution. The Fourier transform of the point spread function, {called} the contrast transfer function~(CTF), is denoted by~$\hat{h}$. The measurement~$y$ represents a \mbox{cryo-EM} micrograph affected by the optical configuration of the microscope through convolution with a point spread function. {The CTF is assumed to be known, and in practice {is} estimated from the data~\cite{heimowitz2020reducing}}.

The goal of this appendix is to derive conditions on~$h$ such that one can recover the autocorrelations of~$M$ from~$y$. Once the autocorrelations of~$M$ were recovered, one can estimate the target image using Algorithm~\ref{alg:image_recovery}. The conditions are presented in the following proposition.
\comm{
\begin{proposition}
\label{prop:ctf}
Reconstructing the first three autocorrelations of~$M$ from~$y$~(\ref{eq:ctf_model}) is possible \mbox{if~$\hat{h}[\vec{k}] \ne 0 \text{ for all } \vec{k} \in \mathbb{Z}^2$}, where~$\hat{h}$ is the CTF.
\end{proposition}}
\begin{proof}
Recall that we first compute the first three autocorrelations of the measurement, as presented in~(\ref{eq:AM1}) -~(\ref{eq:AM3}). For the first-order autocorrelation (the mean) of the measurement~$y$ we have that
\begin{align}
\hat{y}[\vec{0}] = \hat{h}[\vec{0}] \hat{M}[\vec{0}],
\end{align}
where~\mbox{$\hat{y} \in \mathbb{C}^{{N \times N}}$} is the Fourier transform of~$y$, and~\mbox{$\hat{M}~\in~\mathbb{C}^{N \times N}$} is the Fourier transform of~$M$. The reconstruction of the mean of~$M$ is possible from
\begin{equation}
\hat{M}[\vec{0}] = \hat{y}[\vec{0}] / \hat{h}[\vec{0}],
\end{equation}
if~$\hat{h}[\vec{0}] \ne 0$.

For the second-order autocorrelation of~$y$, we consider its Fourier transform---the power spectrum---defined as
\begin{equation}
P_y [\vec{k}] = |\hat{y}[\vec{k}]|^2.
\end{equation}
Hence, we have that for any frequency~$\vec{k} \in \mathbb{Z}^2$, the power spectrum of~$y$ is given by
\begin{equation}
P_y[\vec{k}] = |\hat{h}[\vec{k}]|^2 P_M[\vec{k}],
\end{equation}
and reconstruction of the second-order autocorrelation of~$M$ is possible by
\begin{equation}
P_M[\vec{k}] = P_y[\vec{k}] / |\hat{h}[\vec{k}]|^2,
\end{equation}
if~$\hat{h}[\vec{k}] \ne 0 \text{ for all } \vec{k} \in \mathbb{Z}^2$.

For the third-order autocorrelation of~$y$, we consider its Fourier transform---the bispectrum---defined as
\begin{equation}
B_y[\vec{k}_1, \vec{k}_2] = \hat{y}[\vec{k}_1] \hat{y}[\vec{k}_2]^* \hat{y}[\vec{k}_2 - \vec{k}_1].
\end{equation}
Thus, for any pair of frequencies~$\vec{k}_1, \vec{k}_2 \in \mathbb{Z}^2$, the bispectrum of~$y$ is given by
\begin{equation}
B_y[\vec{k}_1, \vec{k}_2] = \hat{h}[\vec{k}_1] \hat{h}[\vec{k}_2]^* \hat{h}[\vec{k}_2 - \vec{k}_1] B_M[\vec{k}_1, \vec{k}_2],
\end{equation}
and again, reconstruction of the third-order autocorrelation of~$M$ is possible by
\begin{equation}
B_M[\vec{k}_1, \vec{k}_2] = B_y[\vec{k}_1, \vec{k}_2] / (\hat{h}[\vec{k}_1] \hat{h}[\vec{k}_2]^* \hat{h}[\vec{k}_2 - \vec{k}_1]),
\end{equation}
\mbox{if~$\hat{h}[\vec{k}] \ne 0 \text{ for all } \vec{k} \in \mathbb{Z}^2$}. To conclude, the reconstruction of all first three autocorrelations of the measurement~$M$ is possible \mbox{if~$\hat{h}[\vec{k}] \ne 0 \text{ for all } \vec{k} \in \mathbb{Z}^2$}.
\end{proof}

\section{Bandlimited functions on the unit disk}
\label{app:bandlimited}
We follow the analysis and notation of~\cite{marshall2020image}. We assume that~$f$ is supported on the unit disk. We assume that it is bandlimited in the basis of Dirichlet Laplacian eigenfunctions on the unit disk~$D = \{\vec{x} \in \mathbb{R}: |\vec{x}| \le 1\}$, which are solutions to the eigenvalue problem
\begin{equation}
\begin{cases}
-\Delta \psi = \lambda \psi &\text{ in } D\\
\psi = 0 &\text{ on } \partial D,
\end{cases}
\end{equation}
where~$-\Delta = -(\partial_{x x} + \partial_{y y})$ is the Laplacian, and~$\partial D$ is the boundary of the unit disk. In polar coordinates~$(r,\theta)$, these eigenfunctions are of the form
\begin{equation}
\psi_{\nu, q}(r,\theta) = J_\nu\left( \lambda_{\nu, q} r \right) e^{i \nu \theta}, \quad \text{for } r \le 1,
\end{equation}
where~$\nu \in \mathbb{Z}_{\ge 0}$,~$J_\nu$ is the~$\nu$-th order Bessel function of the first kind, and~$\lambda_{\nu, q} > 0$ is the~$q$-th positive root of~$J_\nu$. Recall that~$J_\nu$ is a solution to the differential equation
\begin{equation}
f''(r) +  \frac{1}{r} f'(r) + \left(1 - \frac{\nu^2}{r^2} \right) f(r) = 0.
\end{equation}
Therefore, by writing the Laplacian in polar coordinates we have
\begin{align}
-\Delta \psi_{\nu, q}(r,\theta) &= -\left( \partial_{rr} + \frac{1}{r} \partial_r + \frac{1}{r^2} \partial_{\theta \theta} \right) \psi_{\nu, q}(r,\theta) \nonumber\\&= \lambda_{n,q}^2 \psi_{\nu, q}(r,\theta),
\end{align}
that is,~$\lambda_{\nu, q}^2$ is the eigenvalue associated with~$\psi_{\nu, q}$. Hence, the assumption that~$f$ is bandlimited can be written as
\begin{equation}
f(r, \theta) = \sum_{(\nu, q): \lambda_{\nu, q} \le \lambda} \alpha_{\nu, q} \psi_{\nu, q} (r, \theta), \quad \text{for } r \le 1,
\end{equation}
where~$\lambda$ is the bandlimit frequency, and~$\alpha_{\nu, q}$ are the associated expansion coefficients.

\section{Proof of Proposition~\ref{prop:asd}}
\label{app:arbitrary_spacing_distribution_model}
The second-order autocorrelation of {a} measurement~$M$ at shift~$\vec{\ell}_1=({\ell_1}_y, {\ell_1}_x)$ consists of the following components:
\begin{enumerate}
	\item The second-order autocorrelation between the target image~$F$ and~$\vec{\ell}_1$-shifted~$F$:~$S_2^{\alpha} [\vec{\ell}_1]$ (same as in the \mbox{well-separated} case).

	\item The second-order autocorrelation between~$\vec{\ell}_1$-shifted~$F$ and a neighboring~$F$ at location~$\vec{\ell}_{\text{neigh}}=(i, j)$ such that~${\ell_1}_y - (2n-1) \leq j \leq 2n + {\ell_1}_y - 1$ and~${\ell_1}_x - (2n-1) \leq i \leq 2n + {\ell_1}_x - 1$.

	\item Noise components (same as in the \mbox{well-separated} case).
\end{enumerate}
{Overall}, the term that is added to the model in the arbitrary spacing distribution case is given by
\begin{equation}
\label{eq:A}
A^{\alpha}(\xi) [\vec{\ell}_1] = \sum_{j={\ell_1}_y-(2n-1)}^{2n+{\ell_1}_y-1} \sum_{i={\ell_1}_x-(2n-1)}^{2n+{\ell_1}_x-1} { \xi[i, j] S_{2, \text{pair}}^{\alpha}[(i, j) - \vec{\ell}_1]}.
\end{equation}

The third-order autocorrelation of {a measurement}~$M$ at shifts~$\vec{\ell}_1~=~({\ell_1}_y, {\ell_1}_x)$, ~$\vec{\ell}_2=({\ell_2}_y, {\ell_2}_x)$ consists of the following components:
\begin{enumerate}
	\item The third-order autocorrelation between the target image~$F$,~$\vec{\ell}_1$-shifted~$F$ and~$\vec{\ell}_2$-shifted~$F$:~$S_3^{\alpha}[\vec{\ell}_1, \vec{\ell}_2]$ (same as in the \mbox{well-separated} case).

	\item The third-order autocorrelation between~$\vec{\ell}_1$-shifted~$F$,~$\vec{\ell}_2$-shifted~$F$ and a neighboring~$F$ at location~$\vec{\ell}_{\text{neigh}}=(i, j)$ such that~$\max\{{\ell_1}_y, {\ell_2}_y\}-(2n-1) \leq j \leq 2n+\min\{{\ell_1}_y, {\ell_2}_y\}-1$ and~$\max\{{\ell_1}_x, {\ell_2}_x\}-(2n-1) \leq i \leq 2n+\min\{{\ell_1}_x, {\ell_2}_x\}-1$.

	\item The third-order autocorrelation between~$\vec{\ell}_1$-shifted~$F$, a neighboring~$F$ at location~$\vec{\ell}_{\text{neigh}}=(i, j)$ and~$\vec{\ell}_2$-shifted neighboring~$F$ such that~${\ell_1}_y - (2n-1) \leq j \leq 2n+{\ell_1}_y-{\ell_2}_y-1$ and~${\ell_1}_x - (2n-1) \leq i \leq 2n+{\ell_1}_x-{\ell_2}_x-1$.

	\item The third-order autocorrelation between~$\vec{\ell}_2$-shifted~$F$, a neighboring~$F$ at location~$\vec{\ell}_{\text{neigh}}=(i, j)$ and~$\vec{\ell}_1$-shifted neighboring~$F$ such that~${\ell_2}_y - (2n-1)\leq j \leq 2n+{\ell_2}_y-{\ell_1}_y-1$ and~${\ell_2}_x - (2n-1) \leq i \leq 2n+{\ell_2}_x-{\ell_1}_x-1$.

	\item The third-order autocorrelation between~$\vec{\ell}_1$-shifted~$F$, a neighboring~$F$ at location~$\vec{\ell}_{\text{neigh}_1}=(j_1, i_1)$ and~$\vec{\ell}_2$-shifted another neighboring~$F$ from location~$\vec{\ell}_{\text{neigh}_2}=(j_2, i_2)$ such that~${\ell_1}_y - (2n-1) \leq j_1 \leq 2n + {\ell_1}_y - 1$ and~${\ell_1}_x - (2n-1) \leq i_1 \leq 2n + {\ell_1}_x - 1$, and~$\max\{j_1, {\ell_1}_y\} - (2n-1) - {\ell_2}_y \leq j_2 \leq 2n+\min\{j_1, {\ell_1}_y\} - 1 - {\ell_2}_y$ and~$\max\{i_1, {\ell_1}_x\} - (2n-1) - {\ell_2}_x \leq i_2 \leq 2n+\min\{i_1, {\ell_1}_x\} - 1 - {\ell_2}_x$.

	\item The third-order autocorrelation between~$\vec{\ell}_2$-shifted~$F$, a neighboring~$F$ at location~$\vec{\ell}_{\text{neigh}_1}=(j_1, i_1)$ and~$\vec{\ell}_1$-shifted another neighboring~$F$ from location~$\vec{\ell}_{\text{neigh}_2}=(j_2, i_2)$ such that~${\ell_2}_y - (2n-1) \leq j_1 \leq 2n + {\ell_2}_y - 1$ and~${\ell_2}_x - (2n-1) \leq i_1 \leq 2n + {\ell_2}_x - 1$, and~$\max\{j_1, {\ell_2}_y\} - (2n-1) - {\ell_1}_y \leq j_2 \leq 2n+\min\{j_1, {\ell_2}_y\} - 1 - {\ell_1}_y$ and~$\max\{i_1, {\ell_2}_x\} - (2n-1) - {\ell_1}_x \leq i_2 \leq 2n+\min\{i_1, {\ell_2}_x\} - 1 - {\ell_1}_x$.

	\item Noise components (same as in the \mbox{well-separated} case).
\end{enumerate}

{Overall}, the term that is added to the model in the arbitrary spacing distribution case is given \comm{by~\mbox{$B^{\alpha}(\xi) [\vec{\ell}_1, \vec{\ell}_2] + C^{\alpha}(\zeta)[\vec{\ell}_1, \vec{\ell}_2]$}, where~$B^{\alpha}(\xi) [\vec{\ell}_1, \vec{\ell}_2]$ and~$C^{\alpha}(\zeta)[\vec{\ell}_1, \vec{\ell}_2]$ are given in~(\ref{eq:B}) and~(\ref{eq:C}), respectively}.

\begin{figure*}[!t]
\normalsize
\setcounter{MYtempeqncnt}{\value{equation}}
\setcounter{equation}{61}
\comm{
\begin{align}
\label{eq:B}
B^{\alpha}(\xi)[\vec{\ell}_1, \vec{\ell}_2]  &{}:= \sum_{j=\max\{{\ell_1}_y, {\ell_2}_y\}-(2n-1)}^{2n+\min\{{\ell_1}_y, {\ell_2}_y\}-1} \sum_{i=\max\{{\ell_1}_x, {\ell_2}_x\}-(2n-1)}^{2n+\min\{{\ell_1}_x, {\ell_2}_x\}-1}  \xi [i, j] S_{3, \text{pair}}^{\alpha}[\vec{\ell}_2 - \vec{\ell}_1, (i, j) - \vec{\ell}_1] \nonumber\\\nonumber&{+}\: \sum_{j={\ell_1}_y-{\ell_2}_y - (2n-1)}^{2n+{\ell_1}_y-{\ell_2}_y-1} \sum_{i={\ell_1}_x-{\ell_2}_x - (2n-1)}^{2n+{\ell_1}_x-{\ell_2}_x-1}  \xi[i, j] S_{3, \text{pair}}^{\alpha}[\vec{\ell}_2, \vec{\ell}_1 - (i, j)] \\&{+}\:  \sum_{j={\ell_2}_y-{\ell_1}_y - (2n-1)}^{2n+{\ell_2}_y-{\ell_1}_y-1} \sum_{i={\ell_2}_x-{\ell_1}_x - (2n-1)}^{2n+{\ell_2}_x-{\ell_1}_x-1}  \xi[i, j] S_{3, \text{pair}}^{\alpha}[\vec{\ell}_1, \vec{\ell}_2 - (i, j)]
\end{align}
\begin{align}
\label{eq:C}
C^{\alpha}(\zeta)[\vec{\ell}_1, \vec{\ell}_2] &{}:= \nonumber \sum_{j_1={\ell_1}_y - (2n-1)}^{2n + {\ell_1}_y - 1} \sum_{i_1={\ell_1}_x - (2n-1)}^{2n + {\ell_1}_x - 1}  \sum_{j_2=\max\{j_1, {\ell_1}_y\} - (2n-1) - {\ell_2}_y}^{2n+\min\{j_1, {\ell_1}_y\} - 1 - {\ell_2}_y} \sum_{i_2=\max\{i_1, {\ell_1}_x\} - (2n-1) - {\ell_2}_x}^{2n+\min\{i_1, {\ell_1}_x\} - 1 - {\ell_2}_x}      \zeta[(j_1, i_1,j_2,i_2)]   \times \\\nonumber&\times S_{3, \text{pair}}^{\alpha}[(j_1-{\ell_1}_y, i_1-{\ell_1}_x), (j_2+{\ell_2}_y-{\ell_1}_y, i_2+{\ell_2}_x-{\ell_1}_x)]
\\\nonumber&+ \sum_{j_1={\ell_2}_y - (2n-1)}^{2n + {\ell_2}_y - 1} \sum_{i_1={\ell_2}_x - (2n-1)}^{2n + {\ell_2}_x - 1}  \sum_{j_2=\max\{j_1, {\ell_2}_y\} - (2n-1) - {\ell_1}_y}^{2n+\min\{j_1, {\ell_2}_y\} - 1 - {\ell_1}_y} \sum_{i_2=\max\{i_1, {\ell_2}_x\} - (2n-1) - {\ell_1}_x}^{2n+\min\{i_1, {\ell_2}_x\} - 1 - {\ell_1}_x}     \zeta[(j_1, i_1,j_2,i_2)]  \times \\&\times S_{3, \text{trip}}^{\alpha}[(j_1-{\ell_2}_y, i_1-{\ell_2}_x), (j_2+{\ell_1}_y-{\ell_2}_y, i_2+{\ell_1}_x-{\ell_2}_x)]
\end{align}}
\setcounter{equation}{\value{equation}}
\hrulefill
\vspace*{4pt}
\end{figure*}

\allowdisplaybreaks
\section{{Computing the} discrete Fourier transforms of {the} autocorrelations in Section~\ref{subsec:calc_acs}}
\label{app:deriv}
We look at the discrete Fourier transform (DFT) of~$S_2^{\alpha} $ as defined in~(\ref{eq:S2}),~$\hat{S}_2^{\alpha}: \mathcal{J} \rightarrow \mathbb{C}$:
\begin{align}
\label{eq:S2_k_app}
\hat{S}_2^{\alpha} [\vec{k}_1] &= \sum_{\vec{\ell}_1 \in \mathcal{J}} S_2^{\alpha}[\vec{\ell}_1] e^{-2\pi i (\vec{k}_1 \cdot \vec{\ell}_1)/(4n)} \nonumber\\&= \frac{1}{(2n+1)^2} \sum_{\vec{\ell}_1 \in \mathcal{J}} \Big(\frac{1}{2\pi} \int_0^{2\pi} \sum_{\vec{\ell}\in \mathbb{Z}^2} F_\phi [\vec{\ell}] F_\phi [\vec{\ell} + \vec{\ell}_1] d\phi\Big)\times \nonumber\\ &\times e^{-2\pi i (\vec{k}_1 \cdot \vec{\ell}_1)/(4n)} \nonumber\\&= \frac{1}{(2n + 1)^2} \frac{1}{2\pi}  \int_0^{2\pi} \sum_{\vec{\ell}\in \mathbb{Z}^2} F_\phi [\vec{\ell}] e^{-2\pi i ((-\vec{k}_1)\cdot \vec{\ell})/(4n)} \times \nonumber\\ &\times \sum_{\vec{\ell}_1 \in \mathcal{J}} F_\phi [\vec{\ell} + \vec{\ell}_1] e^{-2\pi i (\vec{k}_1 \cdot [\vec{\ell} + \vec{\ell}_1])/(4n)} d\phi \nonumber\\&= \frac{1}{(2n + 1)^2} \frac{1}{2\pi}  \int_0^{2\pi} \hat{F}_{\phi} [-\vec{k}_1] \hat{F}_{\phi} [\vec{k}_1] d\phi \nonumber\\&= \frac{1}{(2n + 1)^2} \frac{1}{2\pi}  \int_0^{2\pi} |\hat{F}_{\phi} [\vec{k}_1]|^2 d\phi,
\end{align}
since~$F_{\phi} [\vec{\ell}]$ is real-valued. {From~(\ref{eq:rot_F}) we see} that the products~$F_\phi [\vec{\ell}] F_\phi [\vec{\ell} + \vec{\ell}_1]$ that appear in~(\ref{eq:S2}) are bandlimited by~$2\nu_{\text{max}}$ with respect to~$\phi$. Therefore, we can replace the integral over~$\phi$ in~(\ref{eq:S2_k_app}) by a summation over angles sampled at the Nyquist rate, that is,
\begin{equation}
\hat{S}_2^{\alpha}[\vec{k}_1] = \frac{1}{(2n + 1)^2} \frac{1}{4\nu_{\text{max}}} \sum_{\nu = 0}^{4\nu_{\text{max}}-1} |\hat{F}_{\phi_{\nu_2}} [\vec{k}_1]|^2,
\end{equation}
where~$\phi_{\nu_2} := 2\pi\nu/(4\nu_{\text{max}})$. {To proceed, we need the following calculation:}
\begin{align}
\label{eq:DC}
\frac{1}{2\pi}& \int_0^{2\pi} F_\phi [\vec{\ell} + \vec{\ell}_2] d\phi \nonumber\\&= \frac{1}{2\pi} \int_0^{2\pi} \sum_{-\nu_{\max}}^{\nu_{\max}} \Bigg(\sum_{q:\lambda_{\nu, q} \le \lambda} \alpha_{\nu, q} \Psi_{\nu, q}[\vec{\ell} + \vec{\ell}_2]\Bigg) e^{i\nu\phi} d\phi \nonumber\\&= \sum_{-\nu_{\max}}^{\nu_{\max}} \Bigg(\sum_{q:\lambda_{\nu, q} \le \lambda} \alpha_{\nu, q} \Psi_{\nu, q}[\vec{\ell} + \vec{\ell}_2]\Bigg) \frac{1}{2\pi} \int_0^{2\pi}  e^{i\nu\phi} d\phi \nonumber\\&= \sum_{-\nu_{\max}}^{\nu_{\max}} \Bigg(\sum_{q:\lambda_{\nu, q} \le \lambda} \alpha_{\nu, q} \Psi_{\nu, q}[\vec{\ell} + \vec{\ell}_2]\Bigg) \delta[\nu] \nonumber\\&= \sum_{q:\lambda_{0, q} \le \lambda} \alpha_{0, q} \Psi_{0, q}[\vec{\ell} + \vec{\ell}_2].
\end{align}
Now we look  at the DFT of~$S_{2, \text{pair}}^{\alpha}$ as defined in~(\ref{eq:S2_neigh}), \mbox{$\hat{S}_{2, \text{pair}}^{\alpha}: \mathcal{J} \rightarrow \mathbb{C}$}. Using~(\ref{eq:DC}) we have that:
\begin{multline*}
S_{2, \text{pair}}^{\alpha} [\vec{\ell}_1] = \frac{1}{(2n + 1)^2} \sum_{\vec{\ell}\in \mathbb{Z}^2}\Big(\sum_{q:\lambda_{0, q} \le \lambda} \alpha_{0, q} \Psi_{0, q}[\vec{\ell}]\Big) \times \\ \times \Big(\sum_{q:\lambda_{0, q} \le \lambda} \alpha_{0, q} \Psi_{0, q}[\vec{\ell} + \vec{\ell}_1]\Big),
\end{multline*}
and
\begin{align*}
\hat{S}&_{2, \text{pair}}^{\alpha} [\vec{k}_1] = \frac{1}{(2n + 1)^2} \sum_{\vec{\ell}_1 \in \mathcal{J}} S_{2, \text{pair}}^{\alpha} [\vec{\ell}_1] e^{-2\pi i (\vec{k}_1 \cdot \vec{\ell}_1)/(4n)} \\&= \frac{1}{(2n + 1)^2} \sum_{\vec{\ell}_1 \in \mathcal{J}} \sum_{\vec{\ell}\in \mathbb{Z}^2} \Big(\sum_{q:\lambda_{0, q} \le \lambda} \alpha_{0, q} \Psi_{0, q}[\vec{\ell}]\Big) \times \\ &\times \Big(\sum_{q:\lambda_{0, q} \le \lambda} \alpha_{0, q} \Psi_{0, q}[\vec{\ell} + \vec{\ell}_1]\Big) e^{-2\pi i (\vec{k}_1 \cdot \vec{\ell}_1)/(4n)} \\&= \frac{1}{(2n + 1)^2} \Big(\sum_{q:\lambda_{0, q} \le \lambda} \alpha_{0, q} \sum_{\vec{\ell}\in \mathbb{Z}^2} \Psi_{0, q}[\vec{\ell}]e^{-2\pi i (-\vec{k}_1 \cdot \vec{\ell})/(4n)}\Big) \times \\ &\times \Big(\sum_{q:\lambda_{0, q} \le \lambda} \alpha_{0, q} \sum_{\vec{\ell}_1 \in \mathcal{J}} \Psi_{0, q}[\vec{\ell} + \vec{\ell}_1] e^{-2\pi i (\vec{k}_1 \cdot (\vec{\ell} + \vec{\ell}_1))/(4n)}\Big) \\&= \frac{1}{(2n + 1)^2}  \Big(\sum_{q:\lambda_{0, q} \le \lambda} \alpha_{0, q} \hat{\Psi}_{0, q}[-\vec{k}_1]\Big) \Big(\sum_{q:\lambda_{0, q} \le \lambda} \alpha_{0, q} \hat{\Psi}_{0, q}[\vec{k}_1]\Big).
\end{align*}
We look at the DFT of~$S_3^{\alpha}$ as defined in~(\ref{eq:S3}),~$\hat{S}_3^{\alpha}: \mathcal{J} \times \mathcal{J} \rightarrow \mathbb{C}$:
\begin{align}
\label{eq:S3_k_app}
&\hat{S}_3^{\alpha} [\vec{k}_1, \vec{k}_2] = \sum_{\vec{\ell}_1, \vec{\ell}_2 \in \mathcal{J}} S_3^{\alpha}[\vec{\ell}_1, \vec{\ell}_2] e^{-2 \pi i (\vec{k}_1 \cdot \vec{\ell}_1 + \vec{k}_2 \cdot \vec{\ell}_2) / (4n)} \nonumber\\&= \frac{1}{(2n + 1)^2} \sum_{\vec{\ell}_1, \vec{\ell}_2 \in \mathcal{J}} \Big(\frac{1}{2 \pi} \int_0^{2 \pi} \sum_{\vec{\ell} \in \mathbb{Z}^2} F_\phi [\vec{\ell}] \times \nonumber\\& \times F_\phi [\vec{\ell} + \vec{\ell}_1] F_\phi [\vec{\ell} + \vec{\ell}_2] d\phi\Big) e^{-2 \pi i (\vec{k}_1 \cdot \vec{\ell}_1 + \vec{k}_2 \cdot \vec{\ell}_2) / (4n)} \nonumber\\&= \frac{1}{(2n + 1)^2} \frac{1}{2 \pi} \int_0^{2 \pi} \sum_{\vec{\ell} \in \mathbb{Z}^2} F_\phi [\vec{\ell}] e^{-2 \pi i (-\vec{k}_1 - \vec{k}_2) \cdot \vec{\ell} / (4n)} \times \nonumber\\& \times \sum_{\vec{\ell}_1 \in \mathcal{J}}  F_\phi [\vec{\ell} + \vec{\ell}_1] e^{-2 \pi i \vec{k}_1 \cdot (\vec{\ell} + \vec{\ell}_1) / (4n)} \times \nonumber\\& \times \sum_{\vec{\ell}_1 \in \mathcal{J}} F_\phi [\vec{\ell} + \vec{\ell}_2] e^{-2 \pi i (\vec{k}_2 \cdot (\vec{\ell} + \vec{\ell}_2)) / (4n)}  d\phi \nonumber\\&= \frac{1}{(2n + 1)^2}\frac{1}{2\pi} \int_0^{2\pi} \hat{F}_\phi [\vec{k}_1] \hat{F}_\phi [\vec{k}_2] \hat{F}_\phi [- \vec{k}_1 - \vec{k}_2] d\phi.
\end{align}
{From~(\ref{eq:rot_F}) we see} that the products~$F_\phi [\vec{\ell}] F_\phi [\vec{\ell} + \vec{\ell}_1] F_\phi [\vec{\ell} + \vec{\ell}_2]$ that appear in~(\ref{eq:S3}) are bandlimited by~$3\nu_{\text{max}}$ with respect to~$\phi$. Therefore, we can replace the integral over~$\phi$ in~(\ref{eq:S3_k_app}) by a summation over angles sampled at the Nyquist rate, that is,
\begin{multline}
\hat{S}_3^{\alpha}[\vec{k}_1, \vec{k}_2] = \frac{1}{(2n + 1)^2} \frac{1}{6\nu_{\text{max}}} \sum_{\nu = 0}^{6\nu_{\text{max}}-1} \hat{F}_{\phi_{\nu_3}} [\vec{k}_1] \times \\ \times \hat{F}_{\phi_{\nu_3}} [\vec{k}_2] \hat{F}_{\phi_{\nu_3}} [-\vec{k}_1 - \vec{k}_2],
\end{multline}
where~$\phi_{\nu_3} := 2\pi\nu/(6\nu_{\text{max}})$. The DFT of~$S_{3, \text{pair}}^{\alpha}$ as defined in~(\ref{eq:S3_neigh}),~$\hat{S}_{3, \text{pair}}: \mathcal{J} \times \mathcal{J} \rightarrow \mathbb{C}$:
\begin{align*}
&\hat{S}_{3,\text{pair}}^{\alpha}[\vec{k}_1, \vec{k}_2] = \sum_{\vec{\ell}_1, \vec{\ell}_2 \in \mathcal{J}} S_{3,\text{pair}}^{\alpha} [\vec{\ell}_1, \vec{\ell}_2] e^{-2\pi i (\vec{k}_1 \cdot \vec{\ell}_1 + \vec{k}_2 \cdot \vec{\ell}_2)/4n} \\&= \frac{1}{(2n + 1)^2} \sum_{\vec{\ell}_1, \vec{\ell}_2 \in \mathcal{J}} \Big(\frac{1}{2\pi} \int_0^{2\pi} \frac{1}{2\pi} \int_0^{2\pi} \sum_{\vec{\ell} \in \mathbb{Z}^2} F_\phi [\vec{\ell}] F_\phi [\vec{\ell} + \vec{\ell}_1] \times \\ &\times F_\eta [\vec{\ell} + \vec{\ell}_2] d\phi d\eta\Big) e^{-2\pi i (\vec{k}_1 \cdot \vec{\ell}_1 + \vec{k}_2 \cdot \vec{\ell}_2)/4n}  \\&= \frac{1}{(2n + 1)^2}\frac{1}{2\pi} \int_0^{2\pi} \frac{1}{2\pi} \int_0^{2\pi} \sum_{\vec{\ell} \in \mathbb{Z}^2} F_\phi [\vec{\ell}] e^{-2\pi i (- \vec{k}_1 - \vec{k}_2)  \cdot \vec{\ell}/4n} \times \\ &\times \sum_{\vec{\ell}_1 \in \mathcal{J}}  F_\phi [\vec{\ell} + \vec{\ell}_1]  e^{-2\pi i \vec{k}_1 \cdot [\vec{\ell} + \vec{\ell}_1]/4n} \times \\&\times \sum_{\vec{\ell}_2 \in \mathcal{J}}  F_\eta [\vec{\ell} + \vec{\ell}_2]   e^{-2\pi i\vec{k}_2 \cdot [\vec{\ell} + \vec{\ell}_2]/4n} d\phi d\eta \\&= \frac{1}{(2n + 1)^2}\frac{1}{2\pi} \int_0^{2\pi} \frac{1}{2\pi} \int_0^{2\pi} \hat{F}_\phi [\vec{k}_1] \hat{F}_\eta [\vec{k}_2] \hat{F}_\phi [- \vec{k}_1 - \vec{k}_2] d\phi d\eta \\&= \frac{1}{(2n + 1)^2}\frac{1}{2\pi} \int_0^{2\pi} \hat{F}_\phi [\vec{k}_1] \hat{F}_\phi [- \vec{k}_1 - \vec{k}_2] d\phi \frac{1}{2\pi} \int_0^{2\pi} \hat{F}_\eta [\vec{k}_2] d\eta \\&= \frac{1}{(2n + 1)^2}\frac{1}{2\pi} \int_0^{2\pi} \hat{F}_\phi [\vec{k}_1] \hat{F}_\phi [- \vec{k}_1 - \vec{k}_2] d\phi \times \\ &\times \frac{1}{2\pi} \int_0^{2\pi} \sum_{\vec{\ell}_2 \in \mathcal{J}}  F_\eta [\vec{\ell} + \vec{\ell}_2]   e^{-2\pi i\vec{k}_2 \cdot (\vec{\ell} + \vec{\ell}_2)/4n} d\eta \\&= \frac{1}{(2n + 1)^2}\frac{1}{2\pi} \int_0^{2\pi} \hat{F}_\phi [\vec{k}_1] \hat{F}_\phi [- \vec{k}_1 - \vec{k}_2] d\phi \times \\ &\times \sum_{\vec{\ell}_2 \in \mathcal{J}} \Bigg(\sum_{q:\lambda_{0, q} \le \lambda} \alpha_{0, q} \Psi_{0, q}[\vec{\ell} + \vec{\ell}_2]\Bigg) e^{-2\pi i\vec{k}_2 \cdot (\vec{\ell} + \vec{\ell}_2)/4n} \\&= \frac{1}{(2n + 1)^2} \sum_{q:\lambda_{0, q} \le \lambda} \alpha_{0, q} \hat{\Psi}_{0, q}[\vec{k}_2] \frac{1}{2\pi} \int_0^{2\pi} \hat{F}_\phi [\vec{k}_1] \hat{F}_\phi [- \vec{k}_1 - \vec{k}_2] d\phi.
\end{align*}
Using~(\ref{eq:rot_F}) we get that the products~$F_\phi [\vec{\ell}] F_\phi [\vec{\ell} + \vec{\ell}_1]$ that appear in~(\ref{eq:S3_neigh}) are bandlimited by~$2\nu_{\max}$ with respect to~$\phi$. Therefore, we can replace the integral over~$\phi$ above by a summation over angles sampled at the Nyquist rate, that is,
\begin{multline*}
\hat{S}_{3,\text{pair}}^{\alpha}[\vec{k}_1, \vec{k}_2] = \frac{1}{(2n + 1)^2} \sum_{q:\lambda_{0, q} \le \lambda} \alpha_{0, q} \hat{\Psi}_{0, q}[\vec{k}_2] \times \\ \times \frac{1}{4\nu_{\max}}\sum_{\nu=0}^{4\nu_{\max}-1} \hat{F}_{\phi_{\nu_2}} [\vec{k}_1] \hat{F}_{\phi_{\nu_2}} [- \vec{k}_1 - \vec{k}_2],
\end{multline*}
where~$\phi_{\nu_2} = 2\pi\nu/(4\nu_{\max})$. Finally, we compute the DFT of~$S_{3, \text{trip}}^{\alpha}$ as defined in~(\ref{eq:S3_trip}),~$\hat{S}_{3, \text{trip}}^{\alpha}: \mathcal{J} \times \mathcal{J} \rightarrow \mathbb{C}$. Using~(\ref{eq:DC}) we get
\begin{multline*}
S_{3, \text{trip}}^{\alpha}[\vec{\ell}_1, \vec{\ell}_2] = \frac{1}{(2n + 1)^2} \sum_{\vec{\ell} \in \mathbb{Z}^2} \Big(\sum_{q:\lambda_{0, q} \le \lambda} \alpha_{0, q} \Psi_{0, q}[\vec{\ell}] \times \\ \times \sum_{q:\lambda_{0, q} \le \lambda} \alpha_{0, q} \Psi_{0, q}[\vec{\ell} + \vec{\ell}_1] \sum_{q:\lambda_{0, q} \le \lambda} \alpha_{0, q} \Psi_{0, q}[\vec{\ell} + \vec{\ell}_2]\Big),
\end{multline*}
and its DFT:
\begin{align*}
&\hat{S}_{3, \text{trip}}^{\alpha} [\vec{k}_1, \vec{k}_2] = \sum_{\vec{\ell}_1, \vec{\ell}_2 \in \mathcal{J}} S_{3, \text{trip}}^{\alpha}[\vec{\ell}_1, \vec{\ell}_2] e^{-2\pi i (\vec{k}_1 \cdot \vec{\ell}_1 + \vec{k}_2 \cdot \vec{\ell}_2)/4n} \nonumber\\&= \frac{1}{(2n + 1)^2} \sum_{\vec{\ell}_1, \vec{\ell}_2 \in \mathcal{J}} \sum_{\vec{\ell} \in \mathbb{Z}^2} \Big(\sum_{q:\lambda_{0, q} \le \lambda} \alpha_{0, q} \Psi_{0, q}[\vec{\ell}] \times \\ &\times \sum_{q:\lambda_{0, q} \le \lambda} \alpha_{0, q} \Psi_{0, q}[\vec{\ell} + \vec{\ell}_1] \times \nonumber\\& \times \sum_{q:\lambda_{0, q} \le \lambda} \alpha_{0, q} \Psi_{0, q}[\vec{\ell} + \vec{\ell}_2]\Big) e^{-2\pi i (\vec{k}_1 \cdot \vec{\ell}_1 + \vec{k}_2 \cdot \vec{\ell}_2)/4n} \nonumber\\&= \frac{1}{(2n + 1)^2} \Big(\sum_{q:\lambda_{0, q} \le \lambda} \alpha_{0, q}  \sum_{\vec{\ell} \in \mathbb{Z}^2}  \Psi_{0, q}[\vec{\ell}] e^{-2\pi i (- \vec{k}_1 - \vec{k}_2)  \cdot \vec{\ell}/4n}\Big) \times \nonumber\\& \times \Big(\sum_{q:\lambda_{0, q} \le \lambda} \alpha_{0, q} \sum_{\vec{\ell}_1 \in \mathcal{J}} \Psi_{0, q}[\vec{\ell} + \vec{\ell}_1] e^{-2\pi i [\vec{k}_1]  \cdot [\vec{\ell} + \vec{\ell}_1]/4n}\Big) \times \nonumber\\& \times \Big(\sum_{q:\lambda_{0, q} \le \lambda} \sum_{\vec{\ell}_2 \in \mathcal{J}} \alpha_{0, q} \Psi_{0, q}(\vec{x} + \vec{x}_2) e^{-2\pi i [\vec{k}_2]  \cdot (\vec{x} + \vec{x}_2)/4n}\Big) \nonumber\\&= \frac{1}{(2n + 1)^2} \Big(\sum_{q:\lambda_{0, q} \le \lambda} \alpha_{0, q} \hat{\Psi}_{0, q}[\vec{k}_1]\Big) \Big(\sum_{q:\lambda_{0, q} \le \lambda} \alpha_{0, q} \hat{\Psi}_{0, q}[\vec{k}_2]\Big) \times \\ &\times \Big(\sum_{q:\lambda_{0, q} \le \lambda} \alpha_{0, q} \hat{\Psi}_{0, q}[- \vec{k}_1 - \vec{k}_2]\Big).
\end{align*}

Let~$\mathcal{V}$ denote the set of all the pairs~$(\nu, q)$ in the expansion~(\ref{eq:expansion}). We define the column vector~$\omega_{\vec{k}, \phi} \in \mathbb{C}^\mathcal{V}$ by~$(\omega_{\vec{k}, \phi})_{\nu, q} = \hat{\Psi}_{\nu, q} [\vec{k}] e^{i\nu \phi}$, and the column vector~$\alpha \in \mathbb{C}^\mathcal{V}$ by~$(\alpha)_{\nu, q} = \alpha_{\nu, q}$; the latter encodes the parameters {describing} the target image. With this notation,~$\hat{F}_\phi [\vec{k}]$ can be expressed compactly as~$\hat{F}_\phi [\vec{k}] = \alpha^\top \omega_{\vec{k}, \phi}$. In addition, let~$\mathcal{V}_0$ denote the set of all the pairs~$(0, q)$. We define the column vector~ \mbox{${\omega_0}_{\vec{k}} \in \mathbb{C}^{\mathcal{V}_0}$} by~$({\omega_0}_{\vec{k}})_{0,q} = \hat{\Psi}_{0, q} [\vec{k}]$, and the column vector~$\alpha_0 \in \mathbb{C}^{\mathcal{V}_0}$ by~$(\alpha_0)_{0, q} = \alpha_{0, q}$. It follows that
\begin{equation*}
\hat{S}_2^{\alpha} [\vec{k}_1] = \frac{1}{4\nu_{\max} (2n + 1)^2} \sum_{\nu = 0}^{4\nu_{\max}} |{\alpha}^\top \omega_{\vec{k}_1, \phi_{\nu_2}}|^2,
\end{equation*}
\begin{equation*}
\hat{S}_{2, \text{pair}}^{\alpha} [\vec{k}_1] = \frac{1}{(2n + 1)^2} \Big({\alpha_0}^\top {\omega_0}_{-\vec{k}_1}\Big) \Big({\alpha_0}^\top {\omega_0}_{\vec{k}_1}\Big),
\end{equation*}
\begin{multline*}
\hat{S}_3^{\alpha} [\vec{k}_1, \vec{k}_2] = \frac{1}{6\nu_{\max} (2n + 1)^2} \sum_{\nu=0}^{6\nu_{\max}-1} \Big(\alpha^\top \omega_{\vec{k}_1, \phi_{\nu_3}}\Big) \times \\ \times\Big(\alpha^\top \omega_{\vec{k}_2, \phi_{\nu_3}}\Big)  \Big(\alpha^\top \omega_{-\vec{k}_1 - \vec{k}_2, \phi_{\nu_3}}\Big),
\end{multline*}
\begin{multline*}
\hat{S}_{3,\text{pair}}^{\alpha}[\vec{k}_1, \vec{k}_2] = \frac{1}{4\nu_{\max} (2n + 1)^2} {\alpha_0}^\top {\omega_0}_{\vec{k}_2} \times \\ \times \sum_{\nu=0}^{4\nu_{\max}-1} \Big(\alpha^\top \omega_{\vec{k}_1, \phi_{\nu_2}}\Big) \Big(\alpha^\top \omega_{-\vec{k}_1 - \vec{k}_2, \phi_{\nu_2}}\Big),
\end{multline*}
\begin{multline*}
\hat{S}_{3, \text{trip}}^{\alpha} [\vec{k}_1, \vec{k}_2] = \frac{1}{(2n + 1)^2} \Big({\alpha_0}^\top {\omega_0}_{\vec{k}_1}\Big) \Big({\alpha_0}^\top {\omega_0}_{\vec{k}_2}\Big) \Big({\alpha_0}^\top {\omega_0}_{- \vec{k}_1 - \vec{k}_2}\Big),
\end{multline*}
where~$\phi_{\nu_2} := 2\pi\nu/(4\nu_{\text{max}})$,~$\phi_{\nu_3} := 2\pi\nu/(6\nu_{\text{max}})$, and~$\nu_{\text{max}}$ is defined in~(\ref{eq:f_steerable}).

In this form, it is straightforward to compute the gradients {which are required for the optimization problem~\ref{eq:optimization}}, in the frequency domain:
\begin{equation}
\label{eq:gS2}
\nabla_{\alpha} \hat{S}_2^{\alpha} [\vec{k}_1] = \frac{1}{2\nu_{\max}  (2n + 1)^2} \sum_{\nu=0}^{4\nu_{\max}} ({\alpha}^\top \omega_{\vec{k}_1, \phi_{\nu_2}}) \omega_{-\vec{k}_1, \phi_{\nu_2}},
\end{equation}
\begin{align}
\label{eq:gS2_pair}
\nabla_{\alpha} \hat{S}_{2, \text{pair}}^{\alpha} [\vec{k}_1] &= \frac{1}{(2n + 1)^2} \Big(({\alpha}^\top \widetilde{\omega_0}_{\vec{k}_1}) \widetilde{\omega_0}_{-\vec{k}_1} \nonumber \\ &+ ({\alpha}^\top \widetilde{\omega_0}_{-\vec{k}_1}) \widetilde{\omega_0}_{\vec{k}_1} \Big),
\end{align}
\begin{align}
\label{eq:gS3}
\nabla_{\alpha} \hat{S}_3^{\alpha} [\vec{k}_1, \vec{k}_2] &=\sum_{\nu=0}^{6\nu_{\max}-1} \frac{1}{6\nu_{\max} (2n + 1)^2} \times \nonumber \\ &\times  \Bigg(\Big({\alpha}^\top \omega_{\vec{k}_1, \phi_{\nu_3}}\Big) \Big({\alpha}^\top \omega_{\vec{k}_2, \phi_{\nu_3}}\Big) \omega_{-\vec{k}_1 - \vec{k}_2, \phi_{\nu_3}} \nonumber\\\nonumber&+ \Big({\alpha}^\top \omega_{\vec{k}_1, \phi_{\nu_3}}\Big) \Big({\alpha}^\top \omega_{-\vec{k}_1 - \vec{k}_2, \phi_{\nu_3}}\Big) \omega_{\vec{k}_2, \phi_{\nu_3}} \\&+ \Big({\alpha}^\top \omega_{\vec{k}_2, \phi_{\nu_3}}\Big) \Big({\alpha}^\top \omega_{-\vec{k}_1 - \vec{k}_2, \phi_{\nu_3}}\Big) \omega_{\vec{k}_1, \phi_{\nu_3}} \Bigg),
\end{align}
\begin{align}
\label{eq:gS3_pair}
\nabla_{\alpha} \hat{S}_{3,\text{pair}}^{\alpha}[\vec{k}_1, \vec{k}_2] & =\sum_{\nu=0}^{4\nu_{\max}-1} \frac{1}{4\nu_{\max} (2n + 1)^2} \times \nonumber \\ &\times \Bigg(\Big(\alpha^\top \omega_{\vec{k}_1, \phi_{\nu_2}}\Big) \Big(\alpha^\top \widetilde{\omega_0}_{\vec{k}_2}\Big) \omega_{-\vec{k}_1 - \vec{k}_2, \phi_{\nu_2}} \nonumber\\\nonumber&+ \Big(\alpha^\top \omega_{\vec{k}_1, \phi_{\nu_2}}\Big) \Big(\alpha^\top \omega_{-\vec{k}_1 - \vec{k}_2, \phi_{\nu_2}}\Big) \widetilde{\omega_0}_{\vec{k}_2} \\&+ \Big(\alpha^\top \widetilde{\omega_0}_{\vec{k}_2}\Big) \Big(\alpha^\top \omega_{-\vec{k}_1 - \vec{k}_2, \phi_{\nu_2}}\Big) \omega_{\vec{k}_1, \phi_{\nu_2}} \Bigg),
\end{align}
\begin{align}
\label{eq:gS3_trip}
\nabla_{\alpha} \hat{S}_{3, \text{trip}}^{\alpha} [\vec{k}_1, \vec{k}_2] &= \frac{1}{(2n + 1)^2} \Big(({\alpha}^\top \widetilde{\omega_0}_{\vec{k}_1}) ({\alpha}^\top \widetilde{\omega_0}_{\vec{k}_2}) \widetilde{\omega_0}_{- \vec{k}_1 - \vec{k}_2} \nonumber\\&+ ({\alpha}^\top \widetilde{\omega_0}_{\vec{k}_1}) ({\alpha}^\top \widetilde{\omega_0}_{- \vec{k}_1 - \vec{k}_2}) \widetilde{\omega_0}_{\vec{k}_2} \nonumber\\&+ ({\alpha}^\top \widetilde{\omega_0}_{\vec{k}_2}) ({\alpha}^\top \widetilde{\omega_0}_{- \vec{k}_1 - \vec{k}_2}) \widetilde{\omega_0}_{\vec{k}_1} \Big).
\end{align}

\end{document}